\documentclass[%
 aip,
 amsmath,amssymb,
 preprint,%
 showkeys,%
]{revtex4-1}
\usepackage{multirow}
\usepackage{graphicx}% Include figure files
\usepackage{dcolumn}% Align table columns on decimal point
\usepackage{bm}% bold math
\usepackage{mwe}
\usepackage[utf8]{inputenc}
\usepackage[T1]{fontenc}
\usepackage{mathptmx}
\usepackage{mathrsfs}
\usepackage{dcolumn}% Align table columns on decimal point
\usepackage{bm}% bold math
\usepackage{todonotes}
\usepackage{tikz}
\usepackage{xcolor}
\usepackage{nicefrac}
\usepackage{array}
\usepackage{graphics}
\usepackage{amssymb}
\usepackage{amsmath}
\usepackage{graphicx}
\usepackage{multirow}
\usepackage{color}
\usepackage{comment}
\usepackage[normalem]{ulem}
\usepackage{url}
\usepackage{csquotes}
\usepackage{mathrsfs}
\usepackage{tabularx}
\usepackage{subfigure}
\usepackage{float}
\usepackage[makeroom]{cancel}

\newcommand{\rev}[1]{\textcolor{black}{#1}}

\usepackage{calc}
\usepackage{environ}
\usepackage{ragged2e}
\usepackage{etoolbox}

\begin{document}

\title{A Hybrid-DFT Study of Intrinsic Point Defects in $MX_2$ ($M$=Mo, W; $X$=S, Se) Monolayers}

\author{Alaa Akkoush}
\email{alaa.akkoush@gmail.com}
\affiliation{%
 Fritz Haber Institute of the Max Planck Society, Faradayweg 4--6, 14195 Berlin, Germany}%
\affiliation{MPI for the Structure and Dynamics of Matter, Luruper Chaussee 149, 22761 Hamburg, Germany}

\author{\rev{Yair Litman}}
\affiliation{%
 Yusuf Hamied Department of Chemistry,  University of Cambridge,  Lensfield Road,  Cambridge,  CB2 1EW,UK}%

\author{Mariana Rossi}
%\email{Second.Author@institution.edu.}
\email{mariana.rossi@mpsd.mpg.de}
\affiliation{%
 Fritz Haber Institute of the Max Planck Society, Faradayweg 4--6, 14195 Berlin, Germany}%
\affiliation{MPI for the Structure and Dynamics of Matter, Luruper Chaussee 149, 22761 Hamburg, Germany}

\date{\today}% It is always \today, today,

\begin{abstract}
Defects can strongly influence the electronic, optical and mechanical properties of 2D materials, \rev{making} defect stability under different thermodynamic conditions crucial for material-property engineering. In this paper, we present an account of the structural and electronic characteristics of point defects in monolayer transition metal dichalcogenides $MX_2$ with $M$=Mo/W and $X$= S/Se, calculated with density-functional theory using the hybrid HSE06 exchange correlation functional including many-body dispersion corrections. For \rev{the simulation of} charged defects, we employ a  charge compensation \rev{scheme based on the} virtual crystal approximation (VCA). We relate the stability and the electronic structure of charged vacancy defects in monolayer MoS$_2$ \rev{to} an explicit calculation of the S monovacancy in  MoS$_2$ supported on Au(111), and find convincing indication that the defect is negatively charged. Moreover, we show that the \rev{finite-temperature} vibrational contributions to the free energy of defect formation can change the stability transition between adatoms and monovacancies by 300--400 K.  Finally, we probe defect vibrational properties by calculating a tip-enhanced Raman scattering image of \rev{a vibrational mode of a} MoS$_2$ cluster \rev{with and without an S monovacancy}.
\end{abstract}


\maketitle

\section{\label{sec:intro}Introduction}

Transition metal dichalcogenide (TMDC) materials are the subject of intense research, motivated by the possibility of realizing and exploiting novel material properties with ease. The chemical composition of these materials is $MX_2$, where $M$ is a transition metal atom from groups IV-X and $X$ are chalcogenide atoms, which are stacked in $X-M-X$ layered structures in the bulk. The layers are bonded by van der Waals interactions and thus easy to exfoliate or grow as single layers. Semiconductor TMDCs with $M=$Mo, W and $X$=S, Se exhibit an indirect band gap that becomes a direct gap at the Brillouin-zone $K$ point in the monolayer limit, as a consequence of quantum confinement\cite{bandgap_ML,bandgap_ML_2}. In addition, because of the moderate and quasi-2D electronic screening~\cite{ThygesenReview2017}, these materials also present a high exciton binding energy, \rev{resulting in stable} excitons at elevated temperatures\cite{ugeda2014giant}. These characteristics make these materials highly desirable for optoelectronic and many other applications. \cite{WSe2_ML, zeng2012valley,yoon2011good, ParkWang2021}

Defects such as vacancies, intercalation, and substitutional atoms are inevitably present in TMDC monolayers generated by any experimental technique~\cite{LinDefectEng_2016, ding2021quantify} and often also created on purpose. Because it is easy to reach a high concentration of defects in these materials and \rev{therefore} induce \rev{significant} changes in (opto)electronic properties~\cite{tongay2013defects,zhou2013intrinsic,mccreary2016distinct}, \rev{the literature has given much} attention to the characterization of defects in TMDCs \cite{LiangDefectEng2021, TanFreysoldt2020,komsa2015native, KC_2014, komsa_WS2, BertoldoThygesenDatabase2022, komsa_mos2}.
The presence of defects can be detrimental or advantageous, depending on the targeted property. 
To cite a few examples, defect-bound neutral excitons have been shown to form characteristic features in the photoluminescence spectra of monolayer  TMDCs~\cite{chow2015defect} and chalcogen vacancies have been connected to the dynamics of grain boundaries that strongly impact electronic transport properties~\cite{lin2015vacancy}. The presence of defects can also serve as an anchor to dock organic molecules and build robust organic-inorganic interfaces with 2D materials, that allow, for example, the fabrication of field-effect transistor biosensors~\cite{FathiMahjouri2021, LinDefectEng_2016}.

Numerous theoretical studies, \rev{which we discuss throughout this paper}, were \rev{carried out} on these systems. These studies have provided a comprehensive understanding of the stability of intrinsic point defects. Nevertheless, a few important aspects still deserve a closer examination, such as the vibrational contributions to the thermodynamic stability at elevated temperatures with accurate DFT calculations, the impact of including many-body van der Waals corrections in calculations, the charge state of defects on metal-supported TMDCs, and the local vibrational properties related to the presence of defects. 

In this paper, we report \rev{our results} regarding the the thermodynamic stability of neutral and charged point defects in monolayer MoS$_2$, MoSe$_2$, WS$_2$ and WSe$_2$ utilizing DFT with a hybrid exchange correlation functional (HSE06)\cite{HSE06} and employing many-body van der Waals corrections (MBD)\cite{MBD}. We pay particular attention to the vibrational enthalpic and entropic contributions to the defect formation energies at elevated temperatures. For charged defects, we adopt the virtual crystal approximation (VCA) \cite{RichterScheffler2013, RichterThesisTU} scheme to obtain an effective charge compensation \rev{in periodic calculations}. We  present results with this technique, together with an analysis of the electronic structure of the charged systems and a discussion about the charge state of an S vacancy of monolayer MoS$_2$ adsorbed on Au(111). Finally, we report an analysis of the variations in space-resolved Raman scattering signals due to an S vacancy in a MoS$_2$ cluster.

\section{\label{sec:Results} RESULTS AND DISCUSSION}

\begin{figure}[ht!]
   % \centering
 
    \includegraphics[width=0.5\textwidth]{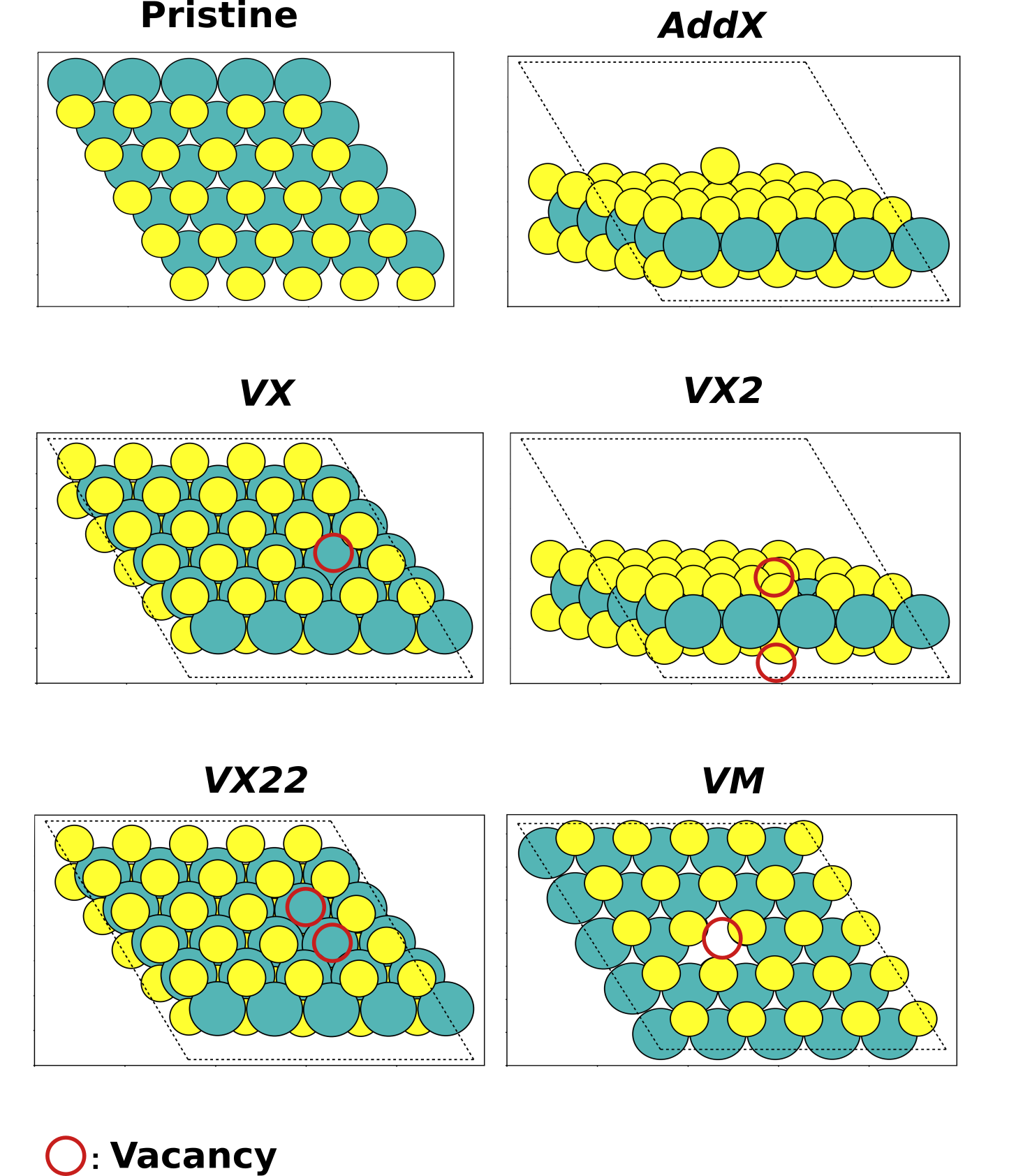}
    \caption{The geometries of the point defects under study for $MX_{2}$, $M$=W, Mo and $X$= Se, S. Add$X$ stands for an $X$ adatom, V$X/M$ stands for $X/M$ monovacancy, V$X2$ stands for $X$ divacancies at the top and bottom coincident lattice sites and V$X22$ stands for $X$ divacancies at neighboring sites. We use these labels to refer to the defects throughout this paper. $M$ atoms are green and $X$ atoms are yellow.}
      \label{fig:Figure1}
\end{figure}

\subsection{Formation Energies of Point Defects}

We have considered monolayer 1H \textit{MX$_2$}, where $M$ stands for Mo, W and $X$ for S, Se. We have investigated the following \rev{common} intrinsic point defects: $X$ monovacancy defects (V$X$);  $M$ monovacancy defects (V$M$); ``up and down''  divacancies (V$X2$), where we removed two $X$ atoms from the top and bottom layers lying on coincident lattice sites; neighboring divacancies (V$X22$), in which two nearest-neighbors $X$ atoms at the same layer are removed; and $X$ adatoms (add$X$), where one $X$ atom is added on top of a host $X$ atom. 
These defects are shown in Fig.~\ref{fig:Figure1}. For V$X$ we have also considered charged defects (+1/-1), as discussed in Section \ref{sec:methods:Formation}.

\begin{figure*}[htbp]
   \includegraphics[width=1\textwidth]{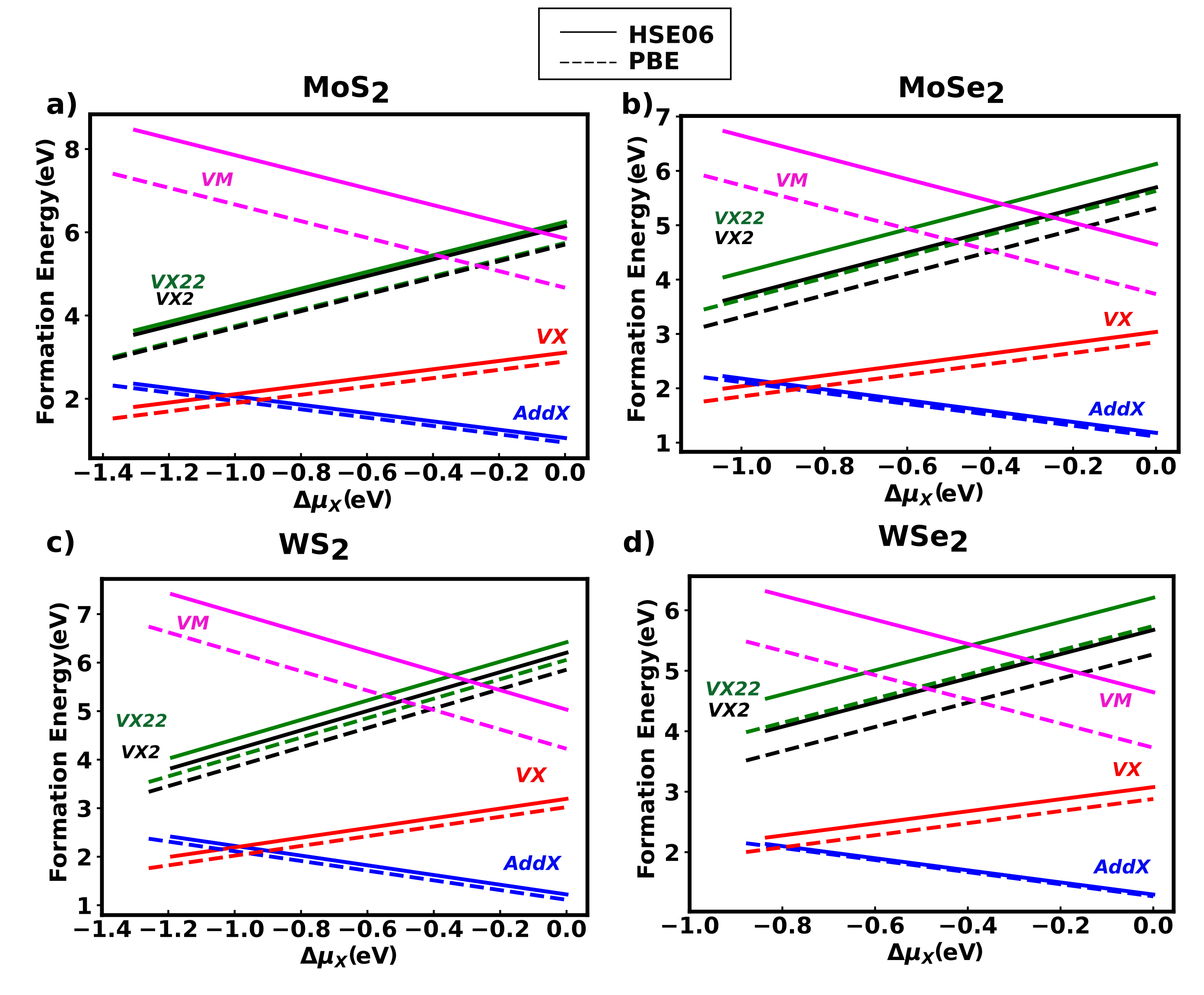}
   \caption{Variation of formation energy (eV) of point defects as a function of $X$ chemical
potential, referenced with respect to the $X$-rich conditions. Dashed lines represent formation energies computed with PBE+MBD and solid lines with HSE06+MBD for a) MoS$_2$, b) MoSe$_2$, c) WS$_2$ and d) WSe$_2$.
\label{fig:Figure2}}
 \end{figure*}

We calculated the formation energies $E^{d}_f$ as in Eq.~\ref{eqn:formation_energy} for the various point defects \rev{shown in Fig.~\ref{fig:Figure1}}, as a function of the possible chemical potentials of $X$=S, Se. \rev{The chemical potentials $\mu_X$ were varied between poor and rich $X$ conditions, as defined in Section~\ref{sec:boundaries}.} We were interested in analyzing the differences between an evaluation of such energies with the PBE+MBD and the HSE06+MBD functionals. These results are shown in Fig.~\ref{fig:Figure2}, \rev{where we referenced $\mu_X$ to the $X$-rich conditions}.
Our results agree with results reported previously in the literature, such as the ones presented in Refs.\cite{KC_2014,guo2020electronic,yang2019electronic, komsa2015native}. 
When improving the description of the electronic structure of these systems, by going from the PBE to the  HSE06 functional, the energetic hierarchy among the various defects remains the same for all systems. 
However, the points at which stability transitions are observed change. In particular, for WSe$_2$ with HSE06 there is no stability transition between add$X$ and V$X$ toward the poor $X$ conditions. We observe the largest differences in formation energies between PBE and HSE06 for the transition-metal vacancies V$M$ in all cases.  This observation could be correlated with differences between PBE and HSE06 predicted band-gaps. Among all defects studied here, the PBE band gaps of V$M$ lie in the range of 0.1 - 0.4 eV, being the smallest band gaps of all defects, as shown in the SI, \rev{Tables S3-S7}. 

The formation energies of add$X$ and V$X$ are always lower than \rev{those} of the other vacancies in either $X$-rich or $M$-rich conditions. Add$X$ appears as the most stable out of all neutral point defects at $X$ rich conditions and over the majority of the possible energy range \rev{of $\mu_X$} (as also reported in Refs.\cite{HaldarSanyal2015, LiHuis2016, KC_2014}). 
As one could expect, the formation energy of divacancies amounts to around twice the formation energy of the monovacancy. However, the results show that for $X$=Se, the \enquote{up \rev{and} down} divacancies VSe2 are more favorable than neighboring VSe22 in all investigated TMDCs (in agreement with Ref.\cite{yang2019electronic}). 

The results presented in this section corroborate most previous work that have investigated defect formation energies in TMDCs \cite{KC_2014,LiHuis2016, komsa2015native,shu2017defect, yang2019electronic}. The consideration of many-body van der Waals effects, absent in most publications in literature, show little impact on these ground-state formation energies.

\subsection{Impact of Temperature and Pressure on Defect Stability}

\begin{figure*}[ht]
   \includegraphics[width=\textwidth]{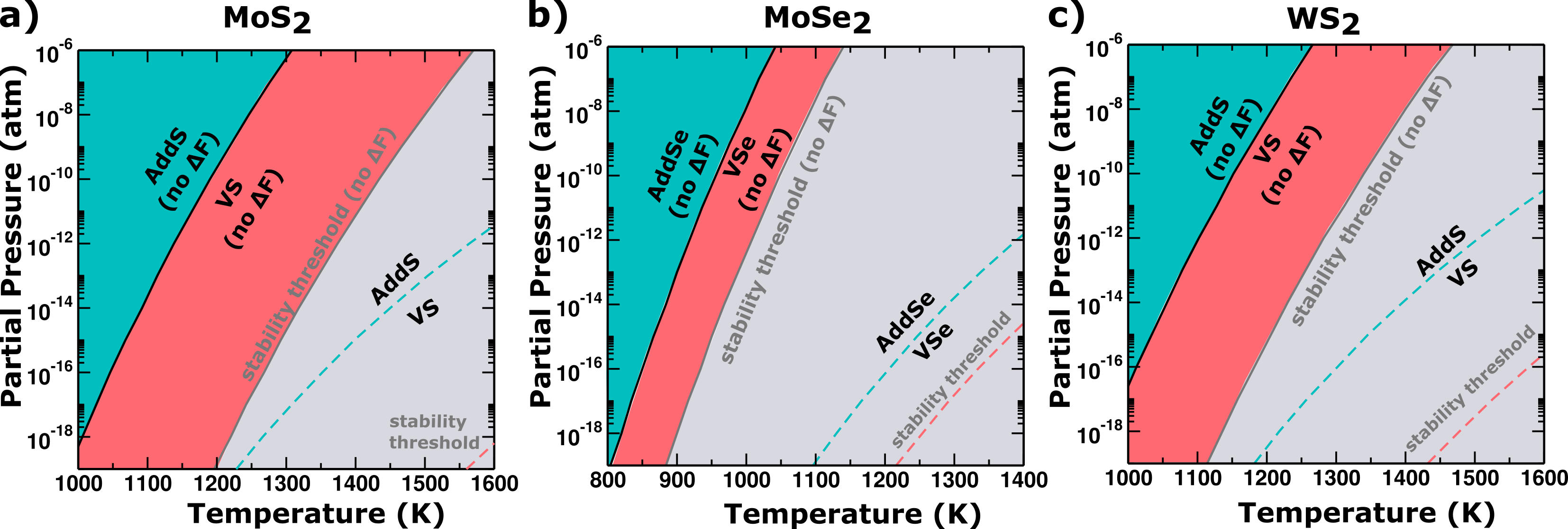}
   \caption{Stability transitions between Add$X$ and V$X$ at different temperatures and partial pressures of S or Se (Eq.~\ref{eqn:total_mu}) for a) MoS$_2$, b) MoSe$_2$ and c) WS$_2$. The full lines represent the boundaries without the vibrational contribution $\Delta F(T)$ and the dashed lines the full formation energy as in Eq.~\ref{eq:Gf_T}.}
        \label{fig:Figure3}
 \end{figure*}

In order to obtain more insights on the defect stability at various thermodynamic conditions, we analysed the connection of the transition points between the most stable defects with temperature and partial pressure. In the calculations, we considered the main contribution of pressure to stem from the chemical potential term, and disregarded lattice expansion effects on the TMDCs. We assume that volume-change contributions will largely cancel when evaluating formation energies. As shown in Ref.~\cite{komsa2015native}, however, at temperatures above 1000 K, the volume changes can amount to differences of $\approx$ 0.2 eV in the formation energies. 

We show in Fig. 
\ref{fig:Figure3} the stability transition lines between V$X$ and Add$X$ for ML  MoS$_2$, MoSe$_2$ and WS$_2$ as a function of temperature and partial S/Se pressure. \rev{We calculate these transitions with and without the temperature-dependent vibrational contributions from the term labeled $\Delta F(T)$ in Eq.~\ref{eq:Gf_T}}. We do not show WSe$_2$ because no stability transition within the boundaries of the chemical potential are predicted for the HSE06+MBD formation energies. We note that considering a different allotrope for the Se reference could \rev{slightly} change this \rev{picture}.  

We first focus on the stability ranges obtained without considering the term labeled $\Delta F(T)$ in Eq.~\ref{eq:Gf_T}. This term is the vibrational Helmholtz free energy difference between the pristine system and the system containing the defect. \rev{This term} is commonly disregarded in these calculations \rev{because it tends to be small in more traditional systems~\cite{RogalReuter2007}}. This means that the $p, T$ dependence of the data represented in Fig.~\ref{fig:Figure3} by the full lines stems only from the terms in Eq.~\ref{eqn:total_mu}\rev{. T}he data presented in Fig.~\ref{fig:Figure2} is \rev{therefore} equivalent to the one presented in Fig.~\ref{fig:Figure3}. However, Fig.~\ref{fig:Figure3} makes it clear that while for MoSe$_2$ the vacancy is stable at much lower temperatures with respect to the S containing systems, its stability range is narrower because the monolayer material ceases to be stable also at lower temperatures when considering equilibrium with the these reservoirs. The stability range of VS on MoS$_2$ and WS$_2$ is larger but starts at higher temperatures. VS in MoS$_2$ \rev{shows} the largest \rev{temperature-}stability range.

\rev{We then} quantify the impact of  $\Delta F(T)$ in the defect formation energy of all materials shown in Fig.~\ref{fig:Figure3}. We observe that including $\Delta F$ (dashed lines in Fig.~\ref{fig:Figure3}) would increase the transition temperature between Add$X$ and V$X$ by 300-400 K, for a given partial pressure. We note that in this case the boundaries of the chemical potential at each temperature are also different because the temperature-dependent vibrational contributions to the bulk and the monolayer must be included in Eq.~\ref{eq:mux-bound}. This naturally raises the question of why  Add$X$ defects are \rev{rarely observed in experiments.}
Since Add$X$ defects are the most stable over a wide range of temperatures and partial pressures, it may be easy to reach larger concentrations of these defects, making it likely that two or more such defects come into contact. For MoS$_2$ it was shown by Komsa and Krasheninnikov~\cite{komsa2015native} that as two AddS defects meet, it becomes favorable to desorb a S$_2$ molecule, especially at elevated temperatures. The increased stability of V$X$ at higher temperatures allied to the proposition that multiple Add$X$ defects can easily desorb could explain why AddS and AddSe are rarely observed in CVD grown TMDCs, while monovacancies are very often observed\cite{ling2014role, wang2014chemical, JinhuaVacancy2015}. 

Therefore, we note that for monolayer TMDCs the vibrational contributions play an important role on the point defect stability. We note that probably this effect is more pronounced due to the high temperature regimes relevant for these systems. At lower temperatures, for example below 600 K, the effect of including or ignoring $\Delta F$ is much less pronounced, as exemplified in Fig. \rev{S4} in the SI.

\subsection{Charged Monovacancies}

\begin{figure}[htbp]
   \includegraphics[width=0.6\textwidth]{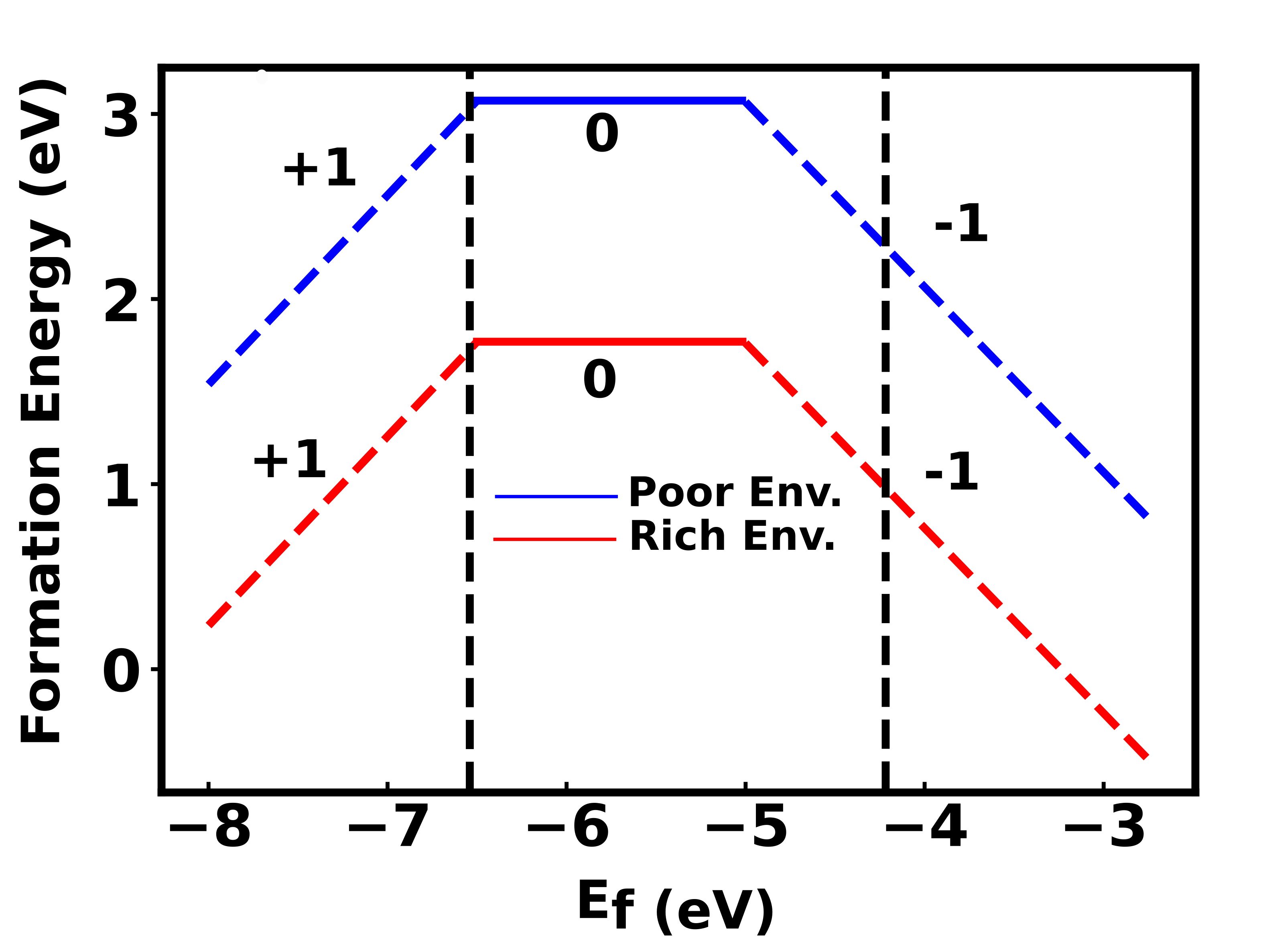}
   \caption{Formation energy of neutral and charged ($q= +1,0,-1 $) VS in MoS$_2$ computed with HSE06+MBD as a function of Fermi-level ($E_f$) in the S-rich (blue) and S-poor (red) conditions. $E_f$ is referenced to the vacuum level. \rev{The dashed lines mark the position of the VBM and the CBM of the pristine MoS$_2$ ML.}}
        \label{fig:Figure4}
 \end{figure}

Next, we \rev{proceeded to analyze} defects that carry an electric charge. Because we have established \rev{that} the qualitative hierarchy of defect formation energies \rev{is similar for} all systems, we focus on the case of MoS$_2$. In addition, we consider only charged S monovacancies (VS), because they are the most \rev{abundant} charged defects appearing in experimentally relevant conditions~\cite{VS_charge,VS_charge_2}. In Fig.~\ref{fig:Figure4} we show the formation energies as calculated from Eq.~\ref{eqn:formation_energy_charged}, with varying $E_f$ and for $\mu_S=0.0$ eV (rich S) and $\mu_S=$-1.3 eV (poor S). We show the data obtained with the charge compensation scheme discussed in Section~\ref{sec:methods:Formation} including corrections to obtain the dilute limit. We note that we performed spin-polarized calculations for the charged defects.  

\rev{In the pristine} MoS$_2$ ML, the computed $E_{\text{VBM}}$ is at -6.54 eV and the $E_{\text{CBM}}$ is at -4.22 eV (HSE06) with respect to the vacuum level. These energies are marked in Fig.~\ref{fig:Figure4}. 
We observe that the positive charge state is predicted to be stable very close to the VBM  (similar to what was reported in Refs.\cite{komsa2015native, TanFreysoldt2020}), while the (0/-1) charge transition level is well within the gap and the negatively charged vacancy is stable for $E_f$ values greater than 1.5 eV above the VBM.

\begin{figure*}[ht]
   \centering
 
    \includegraphics[width=1.0\textwidth]{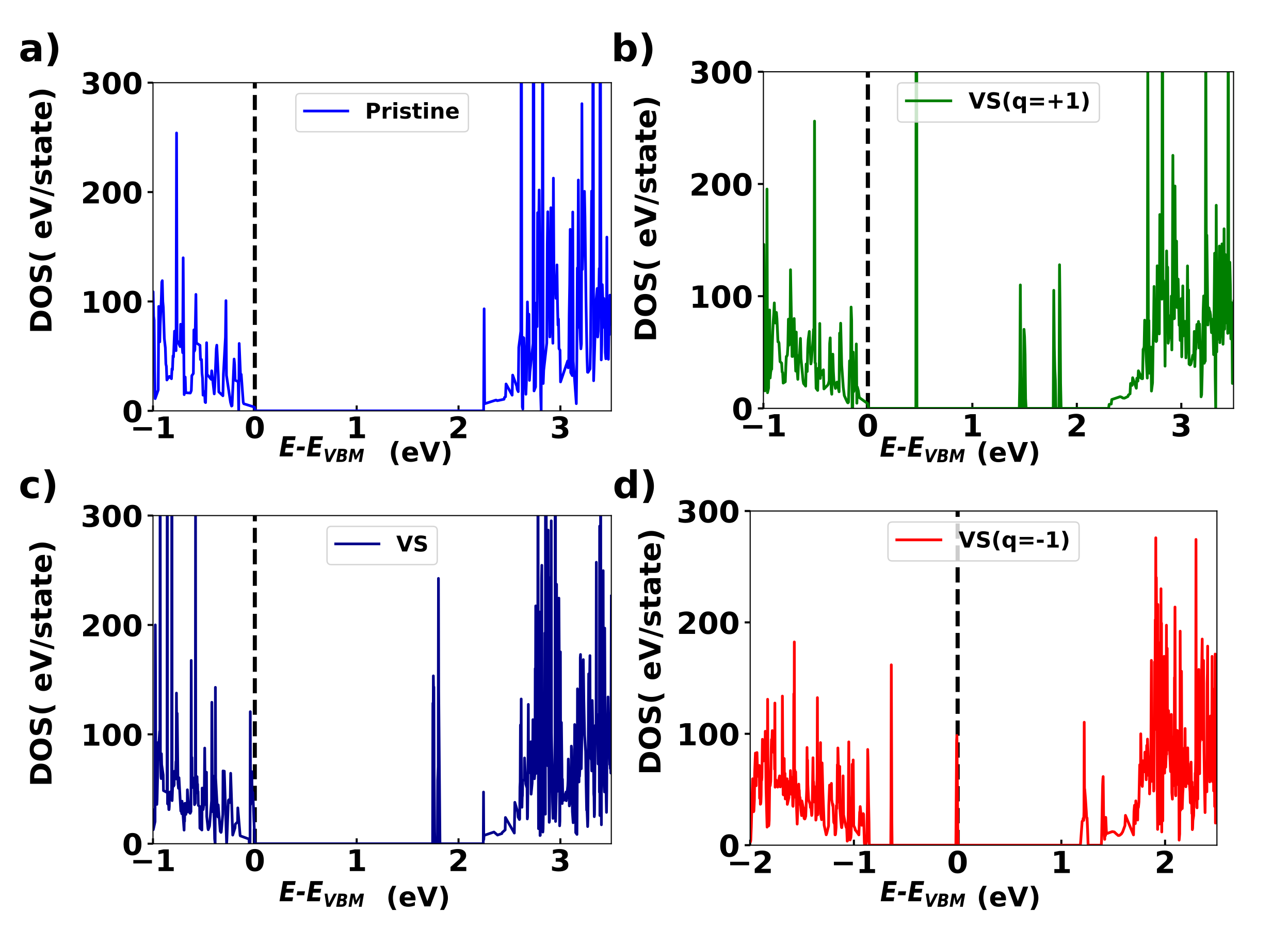}
    \caption{Electronic density of states calculated with the HSE06 functional for (a) pristine MoS$_2$, (b) MoS$_2$ with a positively charged VS (q=+1), (c) MoS$_2$ with a neutral VS and (d) MoS$_2$ with a negatively charged VS (q=-1). }
      \label{fig:Figure5}
\end{figure*}

In Fig.~\ref{fig:Figure5} we compare the electronic density of states (including spin-orbit coupling) of the pristine MoS$_2$ monolayer, the neutral S vacancy, and the charged S vacancies. In all cases, we obtain integer occupation of all energy levels and the ground state of the charged defects is a doublet. 
The results shown for the neutral \rev{VS} confirm DFT results from other authors~\cite{TanFreysoldt2020, zhao2017stability,santosh2014impact}, showing a shallow occupied defect state close to the VBM, and two spin-degenerate unnocupied states in the gap. All these states are of $d$ character and arise from the dangling bonds of the Mo 4$d$ orbitals and the reduced Mo 4$d$ and S 3$p$ orbital hybridization. The splitting between the two unoccupied states is due to spin-orbit coupling. A visualization of the state-resolved electronic density of these defect states is shown in the SI, Fig. \rev{S5}.

We start by discussing the positively charged VS. An unnocupied state with the same character as the shallow occupied state in the neutral VS appears in the gap. This confirms that the orbital that lost one electron is the localized vacancy state, remembering that one spin-channel remains occupied. The vacancy states deep in the gap show a much larger splitting and are not anymore spin-degenerate. As shown in the SI, Fig. \rev{S5}, the states are now grouped by their dominant spin character, and the splitting could be attributed to an exchange interaction with the singly occupied state that lost one electron. We do not observe a structural symmetry breaking around the vacancy. The three Mo atoms around the vacancy form an equilateral triangle with a side length of 3.13 \AA. This is consistent with the fact that all vacancy states show the same character as they had in the neutral case, as shown in Fig. \rev{S5} in the SI. It is worth noting that this geometry is, nevertheless, different from the neutral vacancy, where the equilateral triangle defined by the three neighboring Mo atoms surrounding the vacancy has a side of length 3.04 \AA~in our calculations.

The negatively-charged S vacancy causes a pronounced symmetry breaking on the electronic and atomic structure, characteristic of the Jahn-Teller effect, as discussed previously in Ref.~\cite{TanFreysoldt2020}. Whereas in Ref.~\cite{TanFreysoldt2020} the authors employed GGA and metaGGA functionals, we here corroborate the results with HSE06 and the inclusion of many-body vdW corrections. The structural distortion causes the Mo atoms close to the vacancy to form a isosceles triangle where two sides measure 3.04 \AA~and one side measures 3.16 \AA. The occupied shallow defect state in the neutral case loses its spin degeneracy and one spin channel moves in the gap. The four unnocupied defect states from the neutral vacancy present mixed characters, as shown in Fig. \rev{S5} in the SI, and one of them is fully occupied. The other three (unnocupied) states are found very close to the CBM of the bulk material. 

The calculated stability of the defects and the electronic structure predicted in the calculations are consistent. For completeness, we report the density of states of all neutral point defects under study for MoS$_2$, MoSe$_2$, WS$_2$ and WSe$_2$ computed with HSE06+MBD including SOC in the SI, Figs. \rev{S6--S9}.

\subsection{MoS$_2$ Monovacancy on Au(111)}

\begin{figure}[ht!]
   % \centering
 
    \includegraphics[width=0.48\textwidth]{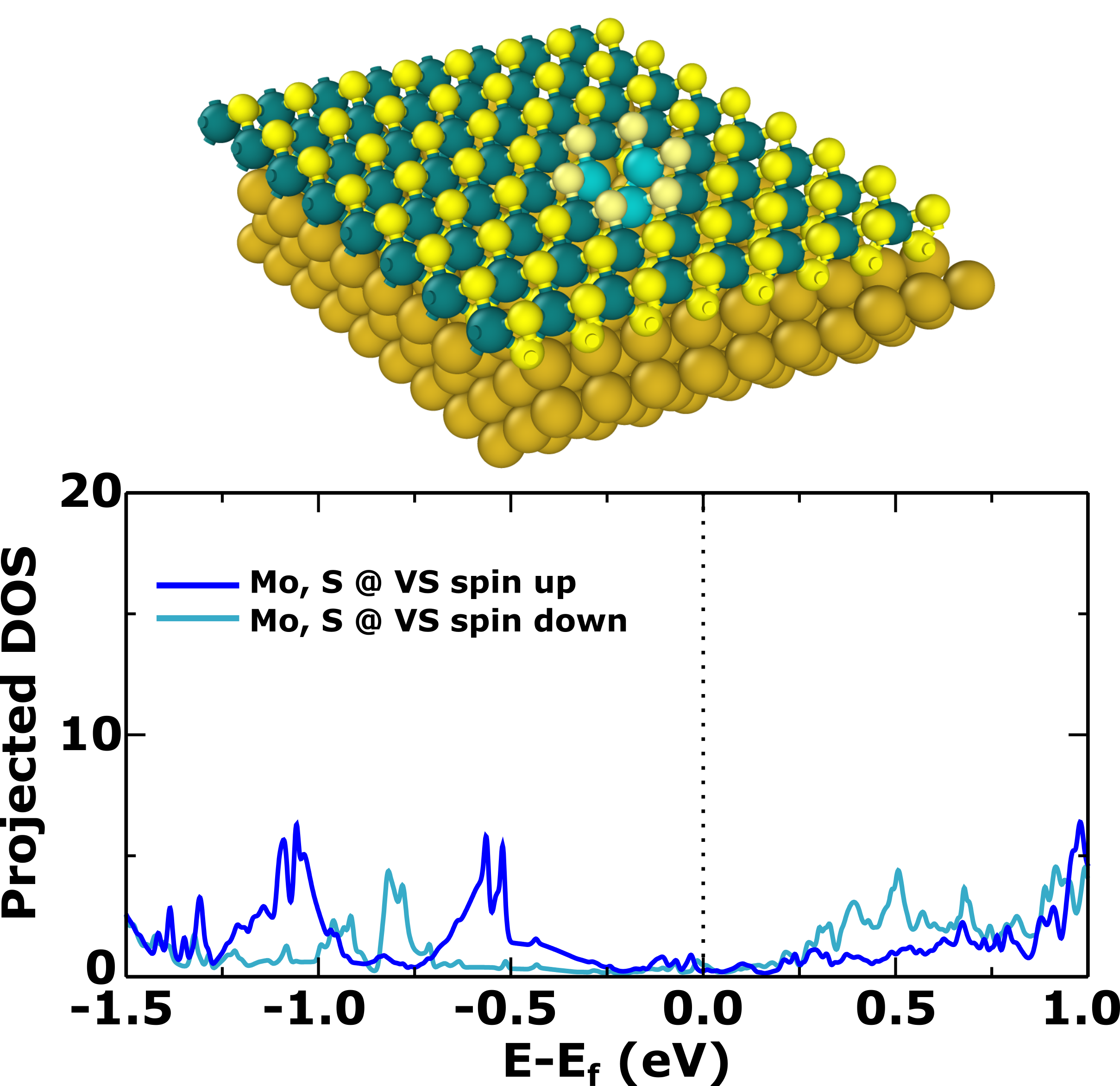}
    \caption{Top: Structure of the $8 \times 8$ supercell of MoS$_2$ on a 4-layer $9 \times 9$ Au(111), where we highlight the Mo and S atoms around the vacancy. Bottom: Projected electronic density of states on the highlighted atoms around the vacancy for spin up and spin down channels. Zero represents the Fermi level of the calculation (defined by the states from the Au surface, not shown), above that the states are unnocupied.}
      \label{fig:Figure6}
\end{figure}

In several situations of interest, MoS$_2$ is supported on a Au(111) substrate~\cite{KraneFranke2018, TuminoTosoni2020}. The bulk Au Fermi energy is calculated to be -4.95 eV with HSE06 and the basis sets used here, and therefore, it could act as a donating electron reservoir that stabilizes a negatively charged vacancy on MoS$_2$. Explicitly simulating the MoS$_2$ monolayer supported by a Au(111) slab requires the use of large supercells in order to minimize the strain induced by the lattice mismatch. We have considered a $8 \times 8$ supercell of MoS$_2$ on a 4-layer $9 \times 9$ Au(111) supercell, containing one S vacancy on the vacuum-facing side of MoS$_2$ (515 atoms), as shown in Fig.~\ref{fig:Figure6}. The MoS$_2$ ML is stretched by 4.8\% in each direction considering HSE06 lattice constants, which induces a small but non-negligible strain on the sheet. However, reducing this number to 1\% would require a $12 \times 12$ supercell of MoS$_2$ on a 4-layer $13 \times 13$ Au(111) surface, at which point the system becomes too large for obtaining results at this level of theory.
We fully relaxed this structure with the HSE06+MBD (same settings as previously in this paper), including spin polarization. We fixed the two bottom Au layers during relaxation. Electronic density of states were calculated with a $4 \times 4 \times 1$ k-point grid for increased accuracy. Fully converging the SCF cycle for this structure with HSE06 functional and the FHI-aims code took around 20 hours when parallelized over 2304 cores in the MPCDF Raven machine (Intel Xeon IceLake-SP 8360Y). We could not apply spin-orbit coupling corrections to this structure with this functional due to technical memory issues.

We observe a Moir\'e pattern formation and a non-uniform distance between the MoS$_2$ layer and Au(111), as also reported in Ref.~\cite{TuminoTosoni2020} where they studied similar systems with the PBE functional and dispersion corrections. In this paper we are interested in understanding whether this vacancy can be considered negatively charged. We confirm that the structure is magnetic and the states with largest spin asymmetries are those of the $d$-orbitals of the Mo atoms around the vacancy. We also observe the tell-tale sign of the pronounced structural distortion around the vacancy, with the Mo atoms forming an isosceles triangle with two sides of 3.36 \AA~and one shorter side of 3.04 \AA. Finally, when analyzing the electronic density projected solely on the atoms surrounding the vacancy, as marked and shown in Fig.~\ref{fig:Figure6}, we observe occupied states that can be assigned to the vacancy at about -0.5 eV, and we confirmed that they are of Mo $d$ character. These states are also singly occupied, as evidenced by the pronounced spin asymmetry between the channels in this region. All of these observations, connected to the discussions in the previous section, point towards a negatively charged vacancy. 

Without further analysis we cannot ascertain the amount of charge at the vacancy. Based on the current data we suppose it is at a -1 charge state. Previous studies that considered a -2 charge state did not find it stable for the monolayer~\cite{komsa2015native}. We note that the structural symmetry breaking is more pronounced and different than the one observed for VS(-1) in free-standing MoS$_2$ in the previous section, which could be due to the structural strain in this case, or indeed a different charge state. We also note that the metallic substrate is known to induce a considerable gap renormalization on the TMDC monolayers due to screening~\cite{Naik2018}. A reduction of the band gap would likely favor the VS(-1) state. Further studies addressing some of these shortcomings and reducing the cost of these large calculations will be the subject of a future work.

\subsection{Local Vibrational Fingerprints}

 \begin{figure}[htb]
   \includegraphics[width=0.7\textwidth]{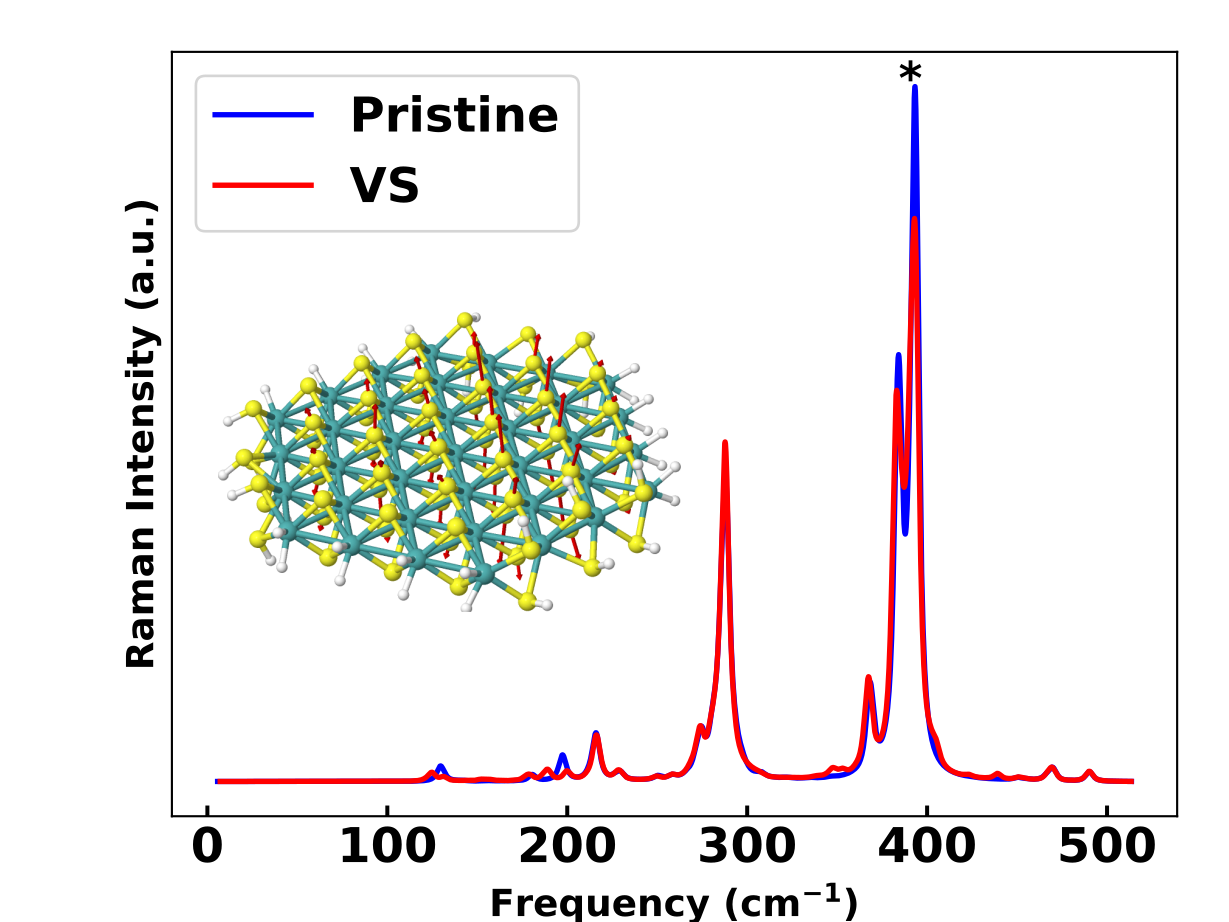}
   \caption{Raman spectra of the MoS$_2$ cluster with and without the sulfur vacancy. The mode shown in the inset is the one marked with an asterisk \rev{for the pristine cluster. The mode resembles the A$_{1g}$ mode of the periodic structure.}}
         \label{fig:Figure7}
  \end{figure}

Raman spectroscopy is a widely used method to characterize the fundamental vibrational properties of 2D materials~\cite{parkin2016raman}. We were interested in \rev{exploring the feasibility of using tip-enhanced  Raman scattering (TERS) signals to obtain a local description of vibrational properties of the vacancies}. The cluster models we use for these calculations do not show the characteristic \rev{Raman active} E$^{1}_{2g}$ (in-plane
 vibrations) and A$_{1g}$ (out-of-plane vibrations) vibrational modes of monolayer MoS$_2$~\cite{parkin2016raman,gupta2022correlated} due to local distortions, but many modes with similar characteristics are present. We show the non-resonant harmonic Raman spectra of the pristine cluster, and the one containing the vacancy in Fig.~\ref{fig:Figure7}. The Raman intensities \rev{shown in Fig.~\ref{fig:Figure7}} were calculated considering \rev{only} the square of the variation of \rev{$\alpha_{zz}$ component of the polarizability tensor} with respect to the normal modes of the system. \rev{The cluster was oriented such that the $z$ axis was perpendicular to the surface plane.}
 Because of the presence of the edges, the clusters show many active Raman modes. The most intense peak for \rev{both systems, lying at 393 cm$^-1$,} corresponds to \rev{a mode that resembles the A$_{1g}$ mode in the periodic structure, and this is the mode that we chose to further characterize by means of a simulation of spatially-resolved TERS.}

\begin{figure}[ht!]
  \centering
   \includegraphics[width=0.49\textwidth]{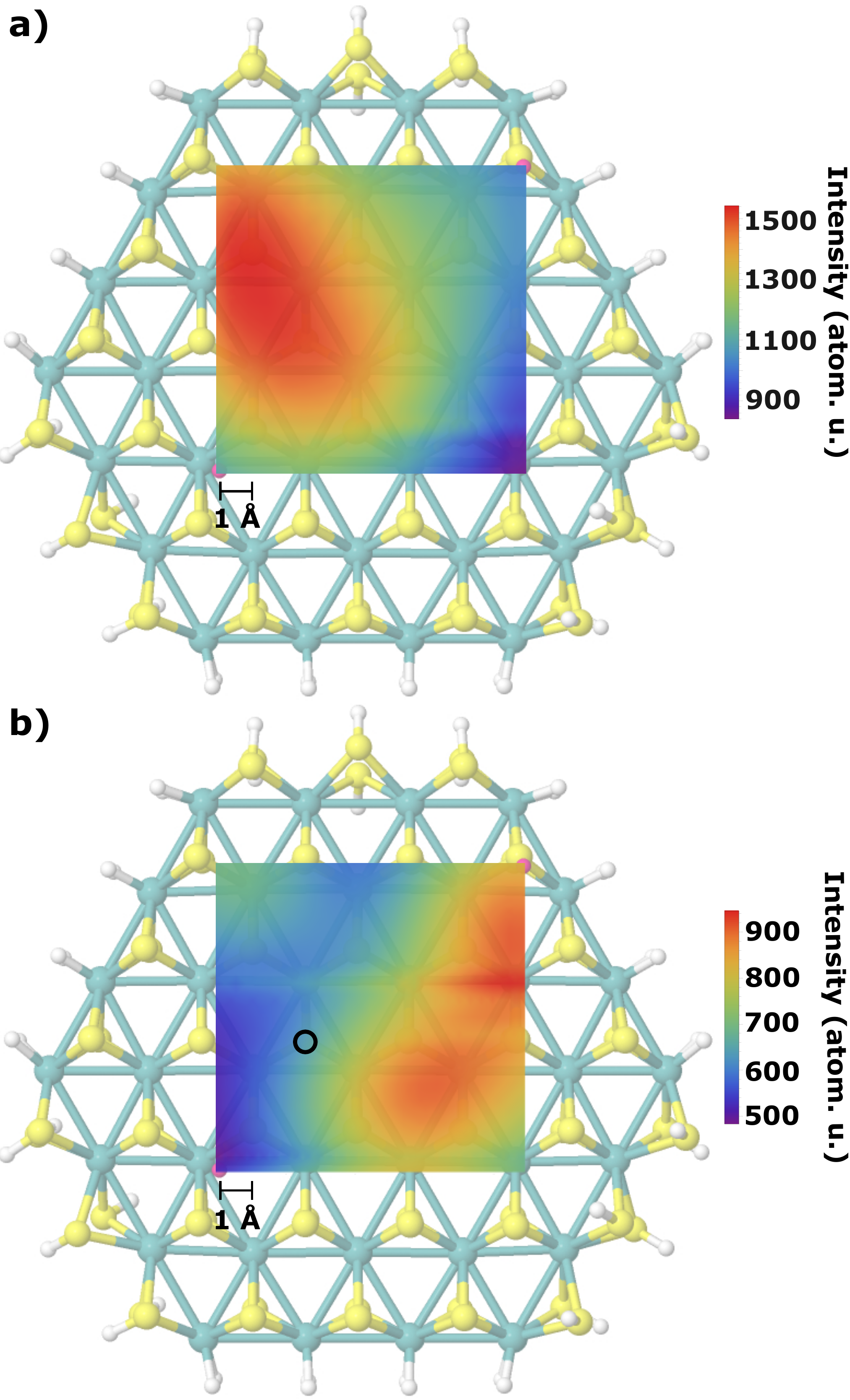}
   \caption{TERS image of \rev{the vibrational modes labeled with an asterisk in Fig. \ref{fig:Figure7}} in the MoS$_2$ cluster corresponding. (a) Pristine system. (b) System including the S vancancy. The position of the S vacancy is marked with a \rev{black} circle \rev{and the pink dots are just visual markers to delimit the image area. Intensities are reported taking into account only the term corresponding to the polarizability variation.}}
        \label{fig:Figure8}
 \end{figure}

\rev{W}e calculated the tip-enhanced Raman intensity according to the methodology proposed in Ref.~\cite{LitmanRossi2022}\rev{, over a region of $9 \times 9$ \AA$^2$ covering} the defect area. These results are presented in Fig.~\ref{fig:Figure8}. We observe that \rev{
 while the Raman spectra of the systems with and without the vacancy shown in Fig.~\ref{fig:Figure7} are very similar, the TERS signals of the pristine system and the vacancy-containing system are substantially different around the vacancy, despite their very similar frequency and overall character. This result shows the possibility of identifying specific vibrational fingerprints of defects in 2D materials even at low defect concentrations.}

 We note that our calculations do not include excitonic states, but these could be included, at least approximately, by performing linear-response time-dependent DFT calculations with an appropriate functional~\cite{Byun_2020}, instead of density-functional perturbation theory calculations within this method. Probably, such a combination would still be considerably more efficient than a full real-time TDDFT calculation of the TERS signal.

\section{\label{sec:Conclusion} Conclusions }

We presented a hybrid DFT study of point defects on semiconductor TMDC monolayers $MX_2$. An analysis of the ground-state formation energy of neutral defects showed that adatom defects are the most stable defects at $X$ rich conditions and through a wide range of chemical potentials. TMDCs containing $X=$S show a small range of S monovacancy stability towards S poor conditions, while this range is reduced for TMDCs containing $X$=Se. A comparison of these formation energies obtained with the PBE+MBD and the HSE06+MBD functionals shows that only quantitative changes in the energy-hierarchy of defect formation energies take place. The largest difference was observed for the Mo and W monovacancies, which could be correlated with the extremely small band gap predicted by PBE+MBD for these systems. Comparing the results obtained in this study and previous results in the literature that did not employ many-body van der Waals corrections, we also conclude that these have a minor quantitative impact on formation energies. This is not surprising, since the main contribution to the defect formation energy in monolayer TMDCs stems from breaking or making covalent bonds. These corrections could have a larger impact in multilayered systems.

\rev{Analyzing the} transitions between V$X$ and Add$X$ \rev{at} temperature \rev{versus} partial pressure \rev{diagrams}, we concluded that VS is the most stable defect only at very elevated temperatures ($>$ 1000 K) for a wide range of partial pressures. VSe is stable at lower temperatures, but its temperature stability range is narrower due to the threshold imposed by the equilibrium with the reservoirs.
We also explicitly quantified the effect of vibrational contributions of the TMDC to the formation free energy. We find that these contributions stabilize Add$X$ defects \rev{and} destabilize V$X$ defects. Disregarding such contributions would lead to a prediction of the stability crossover points between  Add$X$ and V$X$ that would be underestimated by 300--400 K in all materials -- an effect probably exacerbated by the high temperatures at which this transition occurs. These elevated temperatures are nevertheless relevant for some TMDC growth techniques such as chemical vapor deposition.

For charged defects, we find that the virtual crystal approximation (VCA) in an all-electron electronic structure infrastructure is a simple and powerful technique that allows the simulation of charged defects in these 2D systems within a periodic 3D setup. We combined it with a straightforward extrapolation correction for the remaining lateral interactions between charged defects to reach the dilute limit. With this technique, we could confirm the stability of the negatively charged S vacancy in MoS$_2$ with a (0/-1) charge transition level within the gap, and characterized the accompanying Jahn-Teller distortion at the electronic and the atomic structure levels. We then analyzed the electronic and atomic structure of the S vacancy on a MoS$_2$ monolayer supported on Au(111) with the HSE06+MBD functional. This analysis and a comparison to the results of the S vacancy in the free-standing monolayer led us to conclude that the vacancy is negatively charged in this structure. The VCA scheme can be extended to mimic the charge compensation at the Au substrate instead of within the layer and this is the subject of ongoing work. Many of the techniques discussed here could be used in a high-throughput workflow to augment or complement existing data in databases of defects in 2D materials~\cite{BertoldoThygesenDatabase2022}.

In the future, we plan to conduct a deeper analysis of the specific phonon modes that play a role on the stabilization and destabilization of different defects, and their real-space characteristics. In that respect, the exploratory tip-enhanced Raman scattering calculations presented in this work are very encouraging. We believe that a better characterization of the local Raman signal around defects in 2D materials and the possibility of a direct experiment-theory comparison in real space can give unique insights into the atomic motions that accompany charge localization, exciton trapping and polaron formation. We consider such insights particularly interesting to guide the chemical design of organic-inorganic interfaces based on 2D materials for optoelectronic and sensor technologies.

\section{Methods \label{sec:methods}}

\subsection{Basic Parameters \label{sec:methods:parameters}}

Our calculations have been performed using the FHI-aims\cite{blum2009ab} program package and periodic boundary conditions. 
In order to approximate the dilute limit, we aimed at minimizing the interaction between defects in neighboring supercell images. As shown in the SI Fig. S1, the variation of the defect formation energy between a 5$\times$5 supercell and a  7$\times$7 supercell did not exceed 0.03 eV in all studied monolayers. As discussed in the next section, for MoS$_2$ further corrections to the inifinite size limit brought a further 10 meV correction. We thus chose a $5\times5$ supercell to perform most calculations in this work and added a vaccuum region of around 100 \AA~to decouple periodic images in the direction perpendicular to the monolayer surfaces. For charged defects we employ further corrections, as explained in section \ref{sec:methods:Formation}.

\begin{figure}[h!]
   % \centering
   
    \includegraphics[width=0.5\textwidth]{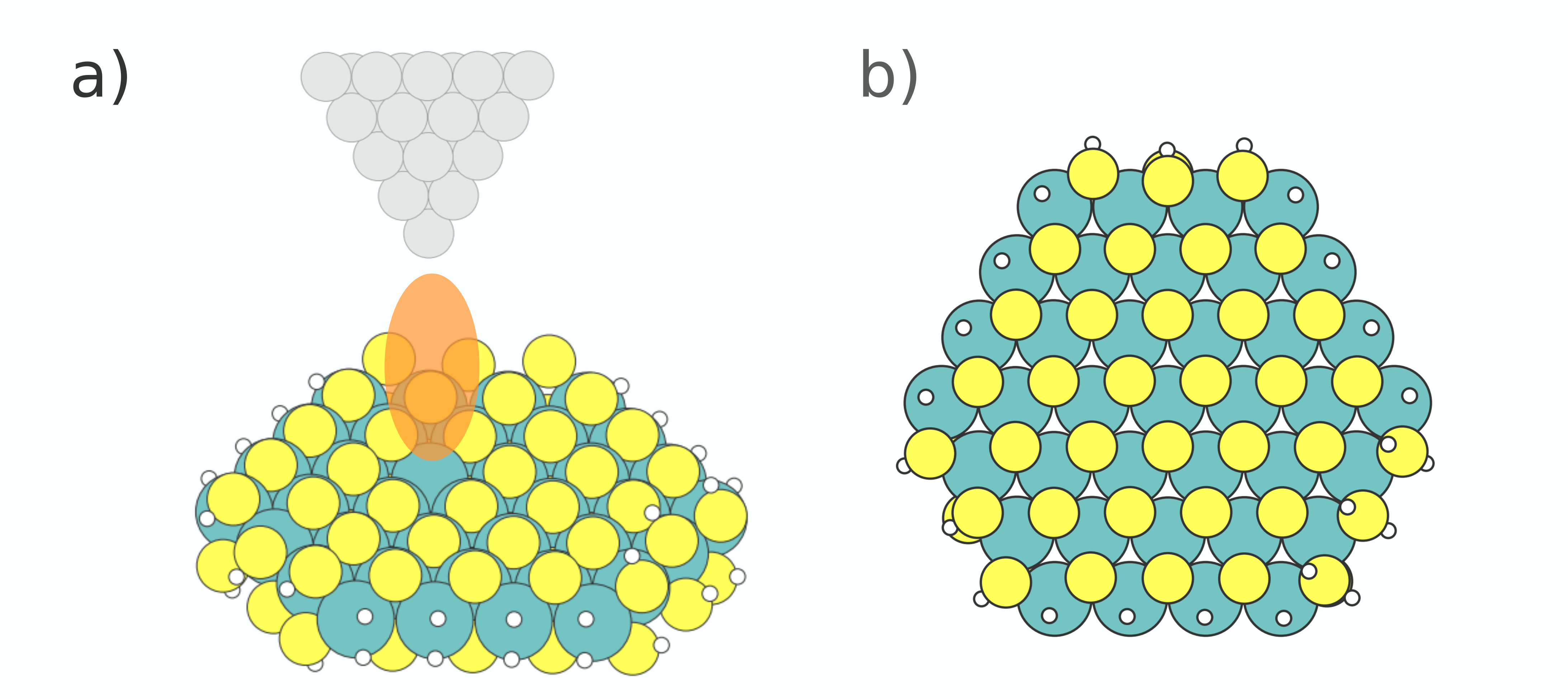}
    \caption{a) Schematic drawing of a TERS setup with an Ag tetrahedral tip over a vacancy defect on an MoS$_2$ cluster. The orange ellipse denotes the local electric field. b) Geometry of the MoS$_2$ cluster used in this study for the calculation of the TERS signals. White circles denote hydrogen atoms.}
    \label{fig:Figure9}
\end{figure}

Electronic structure properties and geometry optimizations were obtained with the Perdew, Burke and Ernzerhof (PBE)\cite{perdew1996generalized} and the HSE06 exchange-correlation functional as proposed by Heyd, Scuseria and Ernzerhof\cite{perdew1996rationale}$^,$\cite{krukau2006influence} with 25\% exact exchange and the screening parameter $\omega$=0.11 Bohr$^{-1}$. Van der Waals (VDW) interactions were accounted for using a
many-body dispersion (MBD) model  \cite{PhysRevLett.124.146401} (HSE06+MBD and PBE+MBD). We performed spin polarized calculations and employed \textit{intermediate} defaults for basis sets and numerical grid settings in the FHI-aims code \cite{levchenko2015hybrid}.  A $k$-grid of 4$\times$4$\times$1 was used for geometry optimizations, total energy evaluations and electronic-structure property calculations. We have included the effect of spin-orbit coupling, known to substantially affect the energy bands of TMDCs~\cite{ZhuChengSchwingen2011}. We employed a ``post-processing" correction, applied only after the electronic ground state density is converged~\cite{PhysRevMaterials.1.033803}. The HSE06+MBD optimized (experimental)  in-plane lattice constants of MoS$_2$, MoSe$_2$, WS$_2$ and WSe$_2$ primitive cells are 3.14 (3.16 \cite{PhysRevB.64.235305}), 3.26 (3.30 \cite{PhysRevB.64.235305}), 3.16 (3.15\cite{1987JSSCh..70..207S}), and 3.27 (3.28\cite{1987JSSCh..70..207S}) \AA~ respectively. 
For phonon calculations, the PBE+MBD functional with ``tight" computational settings in FHI-aims, with the same vdW corrections as described previously (PBE+MBD). These settings were used to obtain vibrational contributions to the formation energy of defects.

Tip-enhanced Raman scattering (TERS) signals were computed using density functional perturbation theory (DFPT) with the LDA functional to compute the density response with respect to a localized electric near-field produced from the response of an Ag tetrahedral tip to an external electric field (Fig.~\ref{fig:Figure9}a). Details of the methodology are presented in reference  \cite{LitmanRossi2022}. Technical problems prevent the near-field response calculation of periodic systems in FHI-aims, thus we employed a cluster approximation. To find a suitable cluster model, we investigated three possible structure forms: hexagonal, rhombic and triangular (details in SI), with different types of hydrogen passivation at the edges. \rev{We fixed the positions of the Mo atoms at the edges of the clusters at their bulk positions}. The choice of the clusters is motivated by studies in the literature such as Refs. \cite{triangularcluster, gronborg2018visualizing}, where MoS$_2$ nanoclusters grown on Au(111) were characterized.
We selected the cluster with the widest band gap upon optimization at the PBE+MBD level. This was a hexagonal cluster of 109 atoms of Mo and S with $\approx$ 43\% S coverage on the edges shown in Fig.~\ref{fig:Figure9}b. Moreover, we confirmed that passivating the cluster edges increased the HOMO-LUMO gaps due to the removal of the dangling bonds, as previously discussed in a DFT study that explored different sizes of MoS$_2$ nanoflakes \cite{javaid2017study}.

\subsection{Defect Formation Energy \label{sec:methods:Formation}}

The formation energy of a defect can provide information about their stability under different thermodynamic conditions. We define this formation energy for neutral defects $E^{d}_{f}$ as follows

\begin{equation}
\label{eqn:formation_energy}
  E^{d}_{f}=E_{d}(n_{i}+\Delta n_{i})-E_{p}(n_{i}) - \sum_{i} \Delta n_{i} \mu_{i},
\end{equation}
where $E_{d}$ and $E_{p}$ are the total energy of the defective system and the pristine system, respectively, $\mu_{i}$ is the chemical potential, and $\Delta n_{i}$ is the number of atoms of type \textit{i} that have been added/removed.

For charged defects, the formation energy gains an extra term
that accounts for the equilibrium with an electron reservoir (Fermi energy),
\begin{equation}
\label{eqn:formation_energy_charged}
  E^{d}_{f}=E_{d}(n_{i}+\Delta n_{i})-E_{p}(n_{i}) - \sum_{i} \Delta n_{i} \mu_{i} +qE_{f}
\end{equation}
where $q$ is the electron charge and $E_{f}$ is effectively an adjustable parameter that depends on the specific conditions under consideration, and is computed relative to the valence band maximum (VBM) of the defect-free system. We only considered charged defects consisting of $X$ vacancies.

In simulations of charged defects in (free standing) 2D materials using supercells with 3D periodic boundary conditions, we encounter the challenge of performing a charge compensation technique that does not generate spurious interactions. If such interactions do arise, they must be corrected~\cite{freysoldt2014first, freysoldt2009fully, Komsa2014charged, HofmannMaurer2021}. The virtual-crystal approximation (VCA)\cite{BellaicheVaderbilt2000, RichterScheffler2013, RichterThesisTU} has been successful in simulating charged defects at bulk systems and surfaces in the past, and removes the spurious interaction of the charged defect with a homogeneous compensating background charge in surface or 2D-material simulations. In this approximation, when working in an all-electron code like FHI-aims, we modify the nuclear charges ($Z'=Z+\Delta Z$) of certain atoms to modify the number of effective electrons in the system, while the simulation as a whole remains neutral. The nuclear charges are modified as the following
 \begin{equation}
\Delta Z =
\left\{
	\begin{array}{ll}
		+\frac{|q|}{N}   & \mbox{for n-type doping} \\
		-\frac{|q|}{N}  & \mbox{for p-type doping}
	\end{array}
\right.
\end{equation}
where $|q|$ is the absolute value of the desired defect charge of the defect and $N$ is the number of atoms for which the nuclear charge is modified. We note that when doping the system in this manner, the reference energies of the pristine and defected systems appearing in the expression of Eq.~\ref{eqn:formation_energy_charged} must be calculated with the same conditions of doping to give a consistent reference. Here we compensate charges within the TMDC monolayer (ML) itself by modifying the nuclear charge of transition metal atoms through the VCA recipe.
We correct the remaining lateral interactions of the charged defects by performing VCA calculations for increasing $L \times L$ supercells with $L=4, 5, 6, 7, 10$ at the PBE+MBD level, fitting a $E_f^d(\infty)+b/L+c/L^2+d/L^3$ function to the formation energies ~\cite{RichterScheffler2013}. This correction is added to the HSE06+MBD formation energies calculated with the $5 \times 5$ supercell and amounts to about \rev{-}30 meV for neutral S vacancies and \rev{+}200 meV for the charged S vacancies \rev{as shown in SI, Fig. S2}.

In all expressions above, the chosen values of the chemical potential $\mu_{i}$ \rev{are} central to the formation energy analysis. We discuss how to model them in the section below. 

\subsubsection{Boundaries of chemical potentials $\mu$ \label{sec:boundaries}}

ML of TMDCs can exist in different conditions of excess of a 
particular constituent atom. In the calculations, we take these possible environments into account by varying the chemical
potential $\mu_i$ between two extremes:
rich $X$ (poor $M$) and poor $X$ (rich $M$) conditions.  

For $MX_{2}$ MLs, one can consider the thermodynamic equilibrium conditions as
\begin{eqnarray}
    \mu_{M}+2\mu_{X}=E^{ML}_{MX_{2}}, \nonumber \\
    \mu_{X}=\frac{1}{2}(E^{ML}_{MX_{2}}-\mu_{M}),
\end{eqnarray}
where $E^{ML}_{MX_{2}}$ refers to the total energy of the primitive unit cell. The lower bound of $\mu_{X}$ takes place for $M$ rich conditions, here modelled by the chemical potential (atomization energy) of $M$ in a bulk BCC structure $\mu_{M}=\mu^{Bulk}_{M}$. With that we obtain
\begin{equation}
  \mu^{min}_{X}=  \frac{1}{2}(E^{ML}_{MX_{2}}-\mu^{Bulk}_{M}).
\end{equation}

The upper bound of $\mu_X$ ($X$ rich environment) is taken as the chemical potential of $X$ in an 8-membered homoatomic ring molecule. This is a common reference in the literature \cite{lehtinen2015atomic,jia2018modulating}. The $S_8$ ring is a predominant S allotrope in the solid and gas-phase~\cite{S8}, while  $Se_8$ is one of three predominately reported Se allotropes in the literature\cite{Se8_1,Se8_2}. 

Therefore,

\begin{equation}
    \mu^{max}_{X}=\frac{1}{8} E_{S_8/Se_8}.
\end{equation}

These considerations lead us to
\begin{equation}
 \mu^{min}_{M}= (E^{ML}_{MX_{2}}-2\mu_X).
\end{equation}

We note that the expressions above automatically determine the boundaries of $\mu_{M}$. 
The final boundaries of chemical potentials that we consider are
\begin{equation}
%\begin{box}
   E^{ML}_{MX_{2}}-2(E_{S_8/Se_8}/8) \leq \mu_{M} \leq \mu^{Bulk}_{M}
%\end{box}
\end{equation}
and 
\begin{equation}
    \frac{1}{2}(E^{ML}_{MX_{2}}-\mu^{Bulk}_{M})\leq \mu_{X} \leq E_{S_8/Se_8}/8.\label{eq:mux-bound}
\end{equation}

\begin{table}[h!]
  \centering
  \caption{Lower boundaries of $\frac{1}{2}\Delta \mu_{M}^{\text{min}}=\Delta \mu_{X}^{\text{min}}=\frac{1}{2}(E^{ML}_{MX_{2}}-\mu^{Bulk}_{M})-E_{X_8}/8$ for the TMDCs under study using HSE06+MBD (PBE+MBD). Values are in eV. }
  \begin{tabular}{c|c|c|c|c}
    \hline
   & MoS$_2$ & MoSe$_2$ & WS$_2$ & WSe$_2$ \\
    \hline
    \hline
    $\Delta \mu_{X}^{\text{min}}$ & -1.30 (-1.37) & -1.04 (-1.09) & -1.19 (-1.26) & -0.83 (-0.88) \\
    \hline
%  $\Delta \mu_{M_i}$ & -1.30 (-1.37) & -1.04 (-1.09) & -1.19 (-1.26) & -0.83 (-0.88) \\
  %  \hline
  \end{tabular}
\end{table}

%%%%%%%%%%%%%%%%%%%%%%%%%%

\subsection{Temperature and pressure contributions to formation energy}

In the following, we consider the temperature and partial pressure contributions on the defect formation energies. 
We consider the Gibbs energy of formation as $G(p,T)=F(V,T)+pV=E-TS+pV$ where $F$ is the Helmholtz free energy, $V$ is the total volume of the system, $p$ stands for pressure, $T$ is the temperature and $S$ is the entropy.
The free energy of defect formation (here considering the case of neutral defects) is given by~\cite{Reuter_2005,freysoldt2014first}
%%%%%%%%%%%%%%%%%%%%%%%% NEW
\begin{equation} \label{eq:gibbs-formation}
  G^{d}_{f}(p,T)=G_{d}(p,T)-G_{p}(p,T) - \sum_{i} \Delta n_{i} \mu_{i}(p,T).
\end{equation}

We consider harmonic vibrational contributions to the Helmholtz vibrational free energy $F(T)$, and a fixed volume. For the reference molecules (chemical potential) we take all vibrational frequencies $\omega_i$ and for the periodic systems we consider those at the $\Gamma$ point of the Brillouin zone of the system supercell. 
Separating this term explicitly in Eq. \ref{eq:gibbs-formation} we obtain
\begin{equation}
 G^{d}_{f}(p,T)= \Delta E+\Delta F(T)-\sum_{i} \Delta n_{i}\mu_{i}(p,T),
 \label{eq:Gf_T}
\end{equation}
where $\Delta E$ is the difference between defect and the pristine ground state total energy and $\Delta F(T)$ is the difference between the respective Helmholtz free energies.
For the chemical potential $\mu_i(T, p)$, we can approximate the partition functions of the rotational, translational and vibrational degrees of freedom of the reference molecular reservoir~\cite{mcquarrie2000statistical}. This leads to the previously reported expressions~\cite{Reuter_2005,freysoldt2014first} 
\begin{multline}
\label{eqn:total_mu}
    \mu(p,T) = \frac{1}{N_{\text{at}}} \bigg\{  -kT\ln \Bigl{[}\left(\frac{2 \pi M }{ h^{2}}\right)^{\frac{3}{2}}\frac{(kT)^{\frac{5}{2}}}{p_{0}}\Bigr{]}-kT \ln\left(\frac{\pi^{\frac{1}{2}}}{\sigma}\right)- \\ 
kT \ln \Bigl{[}\left(\frac{8 \pi kT}{h^{2}}\right)^{\frac{3}{2}}I_{A}^{\frac{1}{2}} I_{B}^{\frac{1}{2}} I_{C}^{\frac{1}{2}}\Bigr{]}+kT \sum_i \ln\Bigl{[}1-\exp \left(-\frac{ \hbar \omega_{i}}{kT} \right)\Bigr{]} \\+ kT \ln\frac{p}{p_{0}} + E_\text{ref} + \sum_{i}\frac{ \hbar \omega_{i}}{2} \bigg\},
\end{multline}
   where $N_{\text{at}}$  is the number of atoms in the molecule or unit cell, $M$ is the total mass of the molecule, $\omega_i$ are the harmonic vibrational frequencies, $p$ is the partial pressure of the species for which the chemical potential is being calculated, $\sigma$ is a molecule-dependent symmetry factor and $I$ are the moments of inertia along the principle axis of rotation of the molecule. The rotational and translational terms are absent for solid-state references. We take $p_0$=1 atm and $E_{\text{ref}}$ as the atomization energy of the chemical species under consideration, for the given standard reference. For S and Se our standard references were the S$_8$ and Se$_8$ molecules in the gas-phase and for Mo and W, the BCC bulk structure. We note that other gas-phase allotropes of S and Se are more stable at elevated temperatures~\cite{S8}.

\begin{acknowledgments}
This work was supported by the Deutsche Forschungsgemeinschaft (DFG) Projektnummer 182087777-SFB 951. We thank Sergey Levchenko for helpful discussions about the VCA corrections and Alan Lewis for helpful discussions about spin states.
\end{acknowledgments}

\bibliography{biblio}% Produces the bibliography via BibTeX.

\end{document}

\title{Supporting Information: A Hybrid-DFT Study of Intrinsic Point Defects in $MX_2$ ($M$=Mo, W; $X$=S, Se) Monolayers}

\author{A. Akkoush}
\email{alaa.akkoush@gmail.com}
\affiliation{%
 Fritz Haber Institute of the Max Planck Society, Faradayweg 4--6, 14195 Berlin, Germany}%
\affiliation{MPI for the Structure and Dynamics of Matter, Luruper Chaussee 149, 22761 Hamburg, Germany}

\author{ Yair Litman}
\affiliation{Yusuf Hamied Department of Chemistry, University of Cambridge, Lensfield Road,
Cambridge, CB2 1EW,UK}
\author{M. Rossi}
\affiliation{%
 Fritz Haber Institute of the Max Planck Society, Faradayweg 4--6, 14195 Berlin, Germany}%
\affiliation{MPI for the Structure and Dynamics of Matter, Luruper Chaussee 149, 22761 Hamburg, Germany}

\date{\today}% It is always \today, today,

\maketitle

\section{\label{sec:results} Simulation Details}
\subsection{Supercell Convergence}
\begin{figure}[ht!]
    \centering
 
    \includegraphics[width=0.7\textwidth]{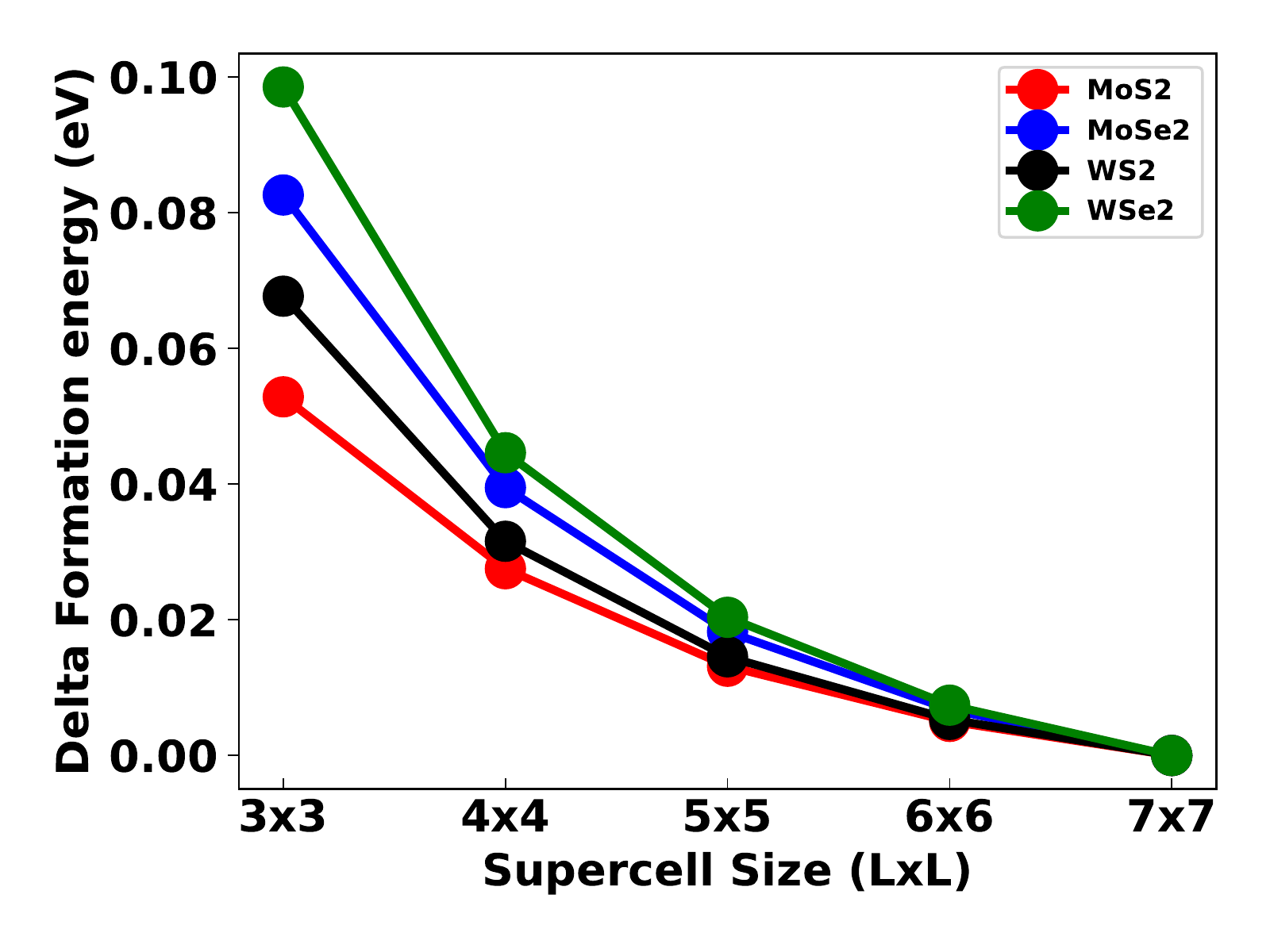}
    \caption{Formation energy convergence of $VX$ in MX$_2$ mono layers with supercell size.}
      \label{fig:convergence}
\end{figure}

In order to identify the appropriate size of the supercell that prevents any potential interaction between defects, we have computed the formation energy difference of $VX$ for the studied TMDC's with respect to various supercell sizes (fig.\ref{fig:convergence}). The calculation is carried out using the PBE+MBD functional.\\

\subsection{Charge Correction}

\begin{figure}[ht!]
    \centering
 
    \includegraphics[width=0.7\textwidth]{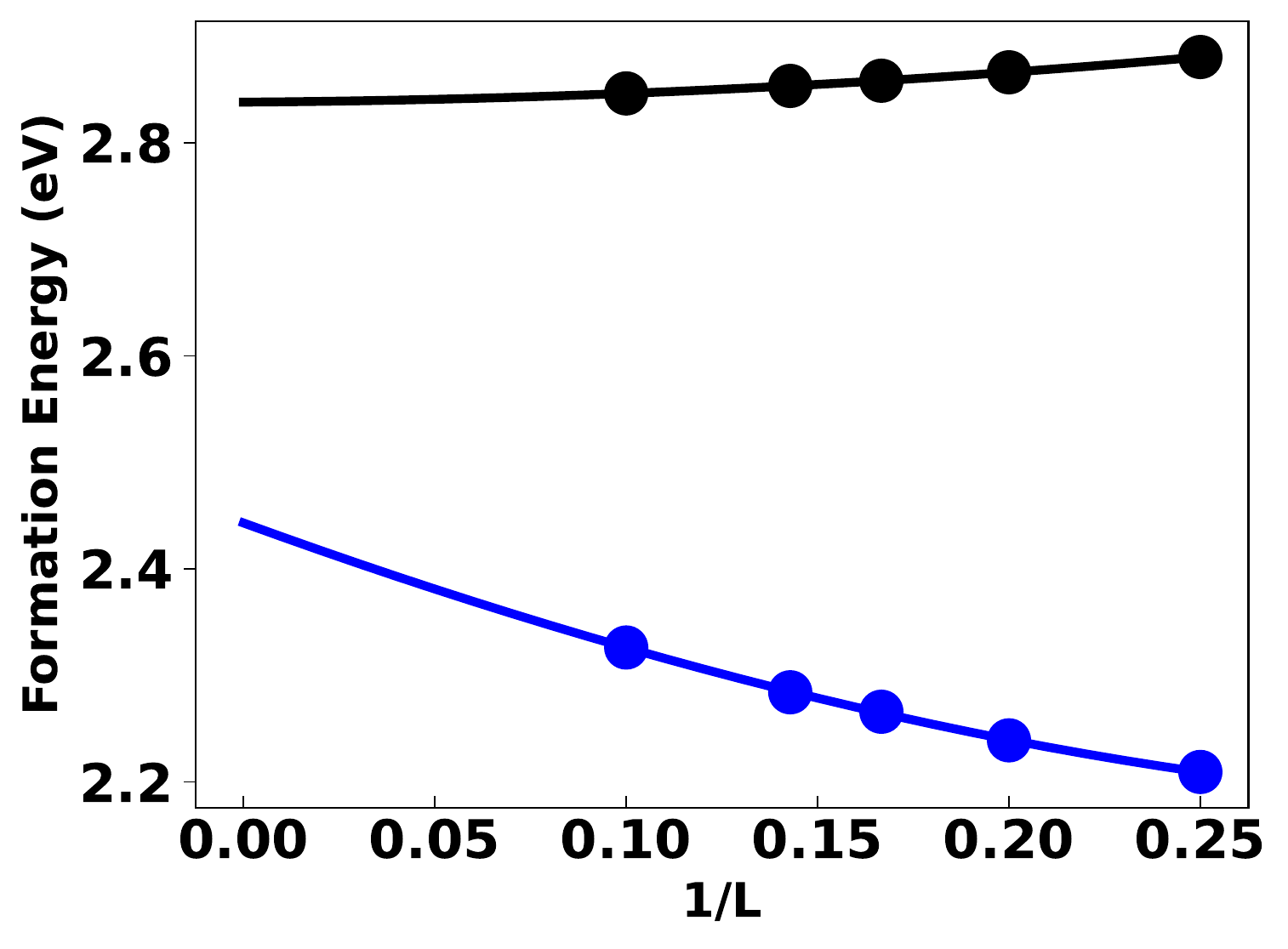}
    \caption{\rev{Formation energy of VS (-1) in blue and neutral VS in black created in MoS$_2$ monolayer using PBE, calculated via the VCA. Solid lines show the defect formation energies extrapolated to the dilute limit of a single defect in an extended material (1/L $\rightarrow$ 0) where L is the multiple of the lattice constant.}}
      \label{fig:convergence_2}
\end{figure}
\newpage
\subsection{Structure Optimization}
\begin{table}[ht!]
  \small
  \centering
  \caption{Geometry Relaxation with HSE06+MBD, a. shows the bond length between $X$ and its corresponding additional $X$ atom on top, b. shows the change between $M-M$ bond lengths  of the equilateral triangle upon the introduction of $X$ vacancy.}
  \subtable[Add$X$]{%
    \hspace{.5cm}%
    \begin{tabular}{c|r|r|r}
        \hline
         MoS$_2$       &  MoSe$_2$    &  WS$_2$ & WSe$_2$ \\
        \hline
        \hline
        1.93       & 2.23      &1.94 & 2.24 \\
        \hline
    \end{tabular}%
    \hspace{.5cm}%
  }
  \hspace{1cm}
  \subtable[V$X$]{%
    \hspace{.5cm}%
    \begin{tabular}{c|r|r|r}
        \hline
       MoS$_2$       &  MoSe$_2$    &  WS$_2$ & WSe$_2$  \\
        \hline
        \hline
        0.10      &    0.17  & 0.15 & 0.22\\
        \hline
    \end{tabular}%
    \hspace{.5cm}%
  }
\end{table}
\subsection{Cluster Search}

\begin{figure}[h!]
    \centering
 
    \includegraphics[width=1.0\textwidth]{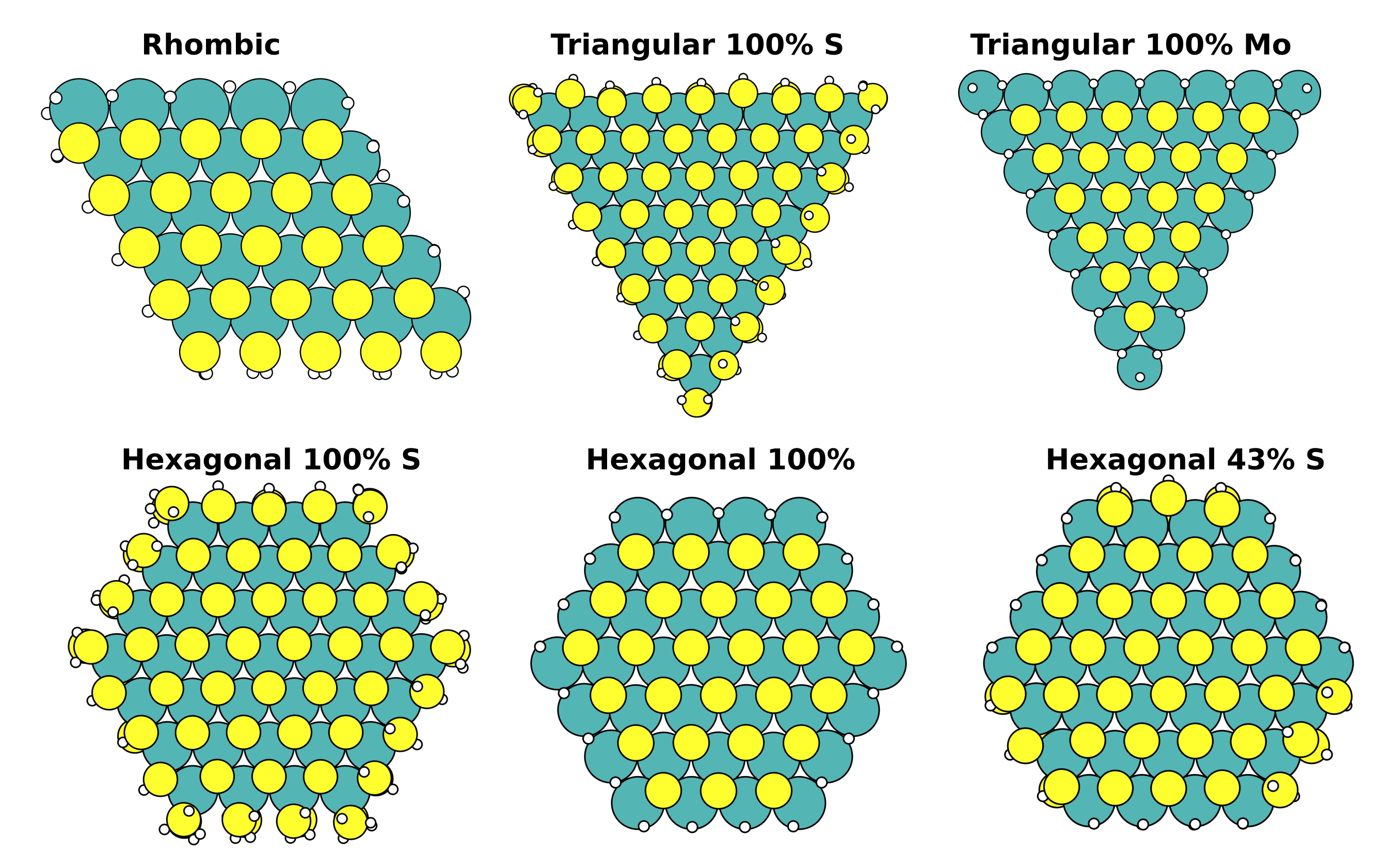}
    \caption{MoS$_2$ cluster structures that has been investigated in this work}
      \label{fig:clusters}
\end{figure}

Different cluster shapes and sizes have been investigated in our search to find an optimal structure. Mainly, we show three types of clusters with different edge atoms (fig.\ref{fig:clusters}). We show in the table S3 the bandgap using HSE06 and PBE for these clusters. In this work we have considered the hexagonal structure with 43\% S edges based on band gap calculations as in table \ref{table:1}. The cluster was passivated by hydrogens to ensure a higher band gap and no edge electron states dominance as shown for the chosen cluster \ref{fig:clusters}.\\

\begin{table}[h!]
    \centering

\begin{tabularx}{\textwidth} { 
  |X|X|X|X|X|X|X| }
 \hline
 Systems & Rhombic 50\% S  & Triangular 100\% S & Triangular 100\% Mo & Hexagonal 100\% S & Hexagonal 43\% S (VS) & Hexagonal 100\% Mo   \\
 \hline
 $E_{gap}$[eV] (PBE)  & 0.139 & 0.045 & 0.024 &   0.014 &0.226 (0.218) &0.0266\\
\hline

$E_{gap}$[eV] (HSE06) &0.093& 0.141 & 0.177  & 0.030 &0.676 (0.662)& 0.274
 \\
\hline
\end{tabularx}

\caption{HOMO-LUMO bandgap of MoS$_2$ clusters using HSE06 and PBE functionals.}
\label{table:1}
\end{table}

\newpage

\subsection{Band Gaps of Point Defects in MX$_2$}
\begin{table}[h!]
    \centering
\begin{tabularx}{\textwidth} { 
  |X|X|X|X|X| }
 \hline
 Systems & MoS$_2$  &  MoSe$_2$ &  WS$_2$ & WSe$_2$    \\
 \hline
 $E_{gap}$[eV] (PBE)  & 0.358 & 0.329 & 0.408 &   0.121 \\
\hline

$E_{gap}$[eV] (HSE06) &0.694& 0.621 & 0.871  & 0.730 
 \\
\hline
\end{tabularx}
\caption{The bandgap variation between HSE06 and PBE of $VM$ point defect in $MX_2$.}
\label{table:2}
\end{table}

\begin{table}[h!]
    \centering
\begin{tabularx}{\textwidth} { 
  |X|X|X|X|X| }
 \hline
 Systems & MoS$_2$  &  MoSe$_2$ &  WS$_2$ & WSe$_2$    \\
 \hline
 $E_{gap}$[eV] (PBE)  & 1.760 & 1.381 & 1.619 &   1.435 \\
\hline

$E_{gap}$[eV] (HSE06) &2.194 & 1.935 & 2.242  & 1.813 
 \\
\hline
\end{tabularx}
\caption{The bandgap variation between HSE06 and PBE of $AddX$ point defect in $MX_2$.}
\label{table:2}

\end{table}

\begin{table}[h!]
    \centering
\begin{tabularx}{\textwidth} { 
  |X|X|X|X|X| }
 \hline
 Systems & MoS$_2$  &  MoSe$_2$ &  WS$_2$ & WSe$_2$    \\
 \hline
 $E_{gap}$[eV] (PBE)  & 1.201 & 1.048& 1.125 &   1.219 \\
\hline

$E_{gap}$[eV] (HSE06) &1.768 & 1.546 & 1.736  &1.475
 \\
\hline
\end{tabularx}
\caption{The bandgap variation between HSE06 and PBE of $VX$ point defect in $MX_2$.}
\label{table:2}
\end{table}
\begin{table}[h!]
    \centering
\begin{tabularx}{\textwidth} { 
  |X|X|X|X|X| }
 \hline
 Systems & MoS$_2$  &  MoSe$_2$ &  WS$_2$ & WSe$_2$    \\
 \hline
 $E_{gap}$[eV] (PBE)  & 1.151 & 1.069& 1.017 &   0.836 \\
\hline

$E_{gap}$[eV] (HSE06) &1.688& 1.512 & 1.603  &1.374
 \\
\hline
\end{tabularx}
\caption{The bandgap variation between HSE06 and PBE of $VX2$ point defect in $MX_2$.}
\label{table:2}
\end{table}
\begin{table}[h!]
    \centering
\begin{tabularx}{\textwidth} { 
  |X|X|X|X|X| }
 \hline
 Systems & MoS$_2$  &  MoSe$_2$ &  WS$_2$ & WSe$_2$    \\
 \hline
 $E_{gap}$[eV] (PBE)  & 0.963 & 0.908& 1.200 &   1.068 \\
\hline

$E_{gap}$[eV] (HSE06) &1.525& 1.458 & 1.675  &1.440
 \\
\hline
\end{tabularx}
\caption{The bandgap variation between HSE06 and PBE of $VX22$ point defect in $MX_2$.}
\label{table:2}
\end{table}
\clearpage
\subsection{Impact of Temperature on the Formation Energy of Defects}
\begin{figure*}[ht]
   \includegraphics[width=1\textwidth]{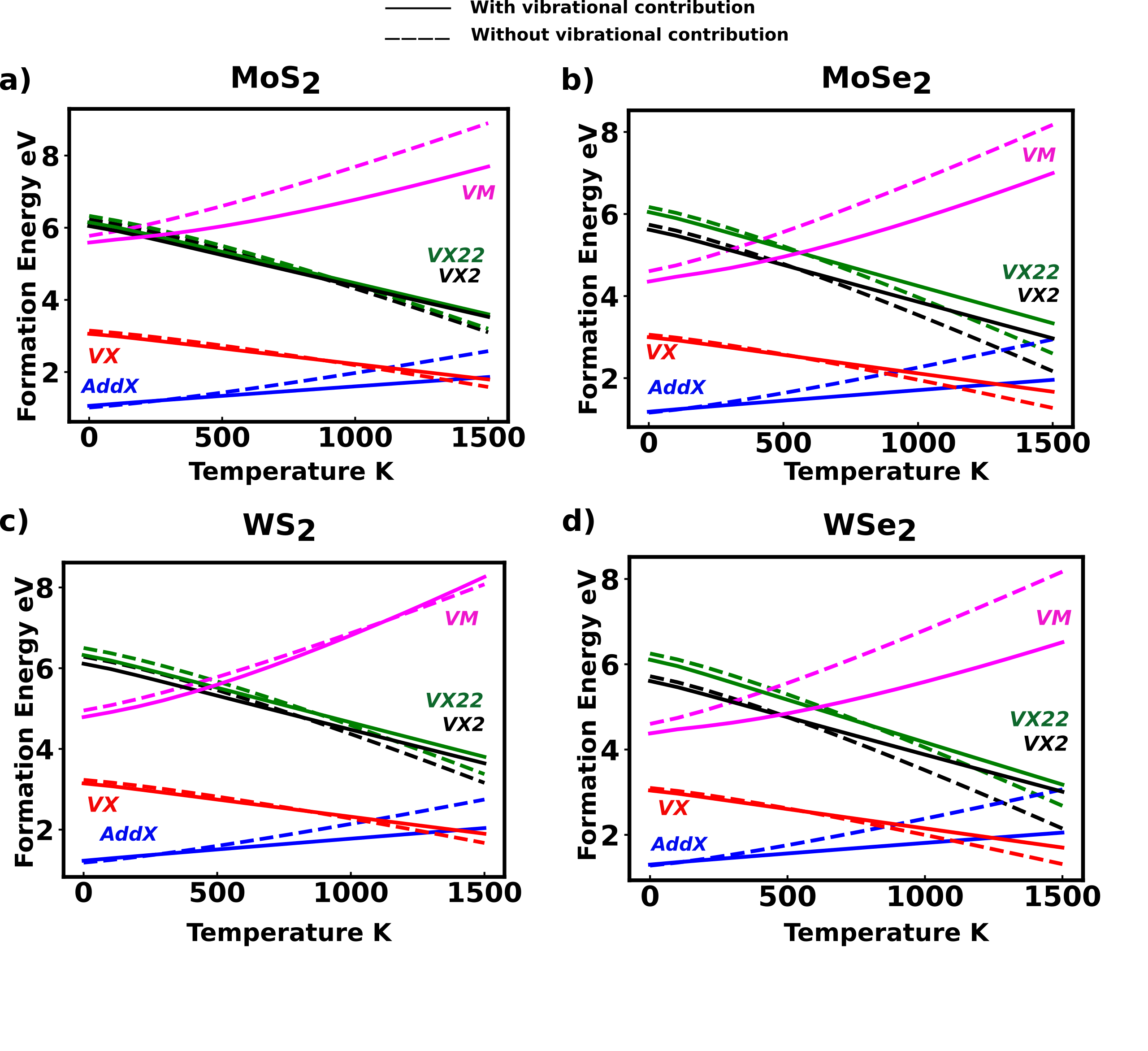}
   \caption{Variation of formation energy (eV) of point defects for $MX_2$ monolayers as a function of temperature calculated with the HSE06+MBD functional at a S and Se partial pressure $p=10^{-14}$ atm. The dashed lines represent the formation energy without the vibrational contribution $\Delta F(T)$ of the TMDCs and the solid lines the full formation energy.}
        \label{Ef_T}
\end{figure*}        
\clearpage
\subsection{Density of States of S Monovacancy in MoS$_2$ at Different Charge States}

\begin{figure}[htbp]
    \centering
 
    \includegraphics[width=\textwidth]{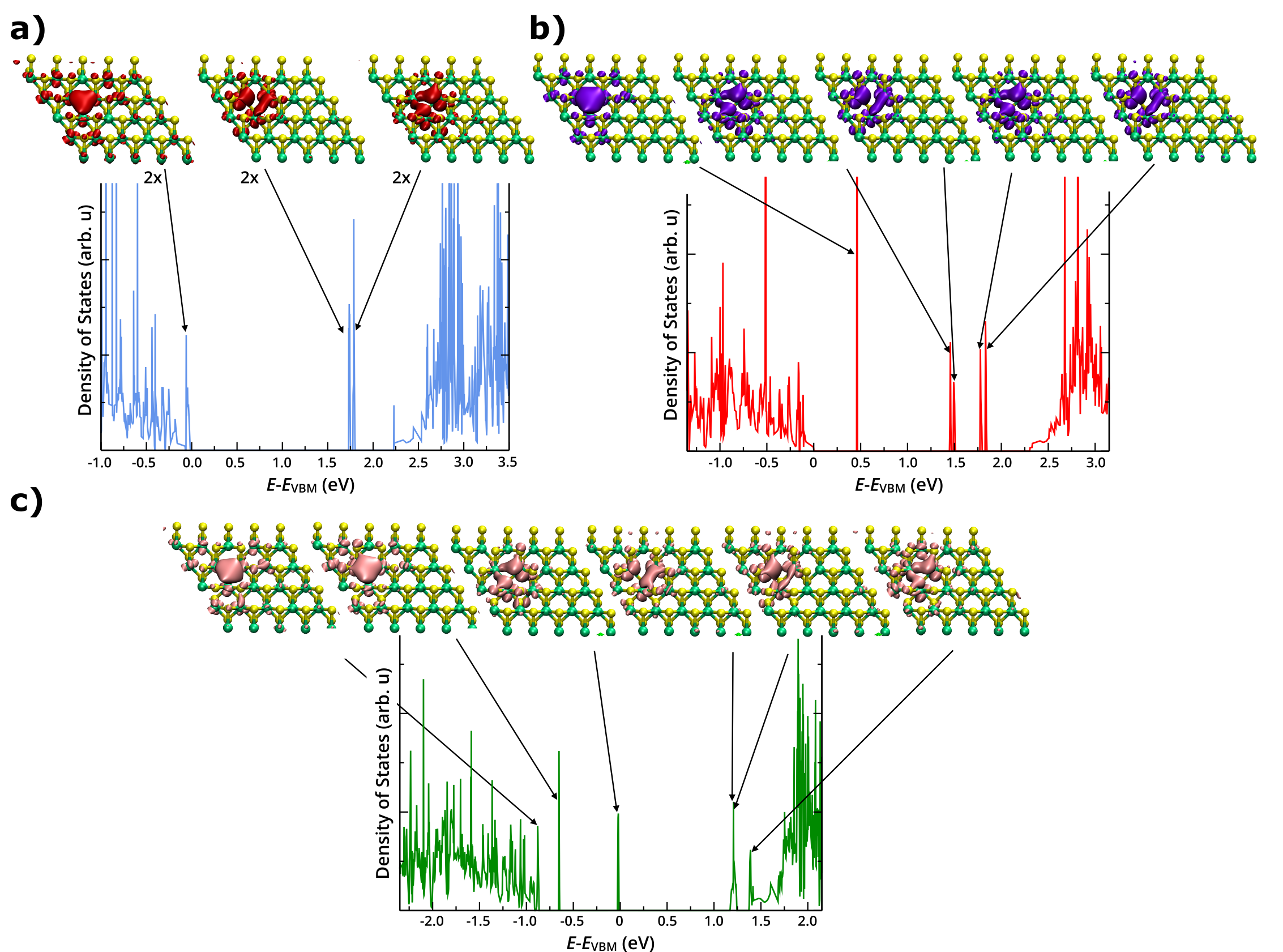}
    \caption{Density of states and state-resolved electronic density of an MoS$_2$ monolayer containing (a) a neutral S monovacany, (b) a positively charged S monovacancy and (c) a negatively charged S monovacancy,  employing the HSE06 functional and including SOC.}
      \label{fig:neutral}
\end{figure}

% \begin{figure}[htbp]
%     \centering
 
%     \includegraphics[width=1.1\textwidth]{pos-state-vis.pdf}
%     \caption{Density of states and electron density of defect states of MoS$_2$ monolayer for positively charged S monovacany using HSE06 exchange correlation, MBD and SOC.}
%       \label{fig:pos}
% \end{figure}
% \begin{figure}[htbp!]
%     \centering
 
%     \includegraphics[width=1.1\textwidth]{neg-state-vis.pdf}
%     \caption{Density of states and electron density of defect states of MoS$_2$ monolayer for negatively charged S monovacany using HSE06 exchange correlation, MBD and SOC.}
%       \label{fig:neg}
% \end{figure}
\newpage

\subsection{Density of States of Point Defects in MX$_2$ (HSE06)}
\begin{figure}[h!]
    \centering
 
    \includegraphics[width=1.0\textwidth]{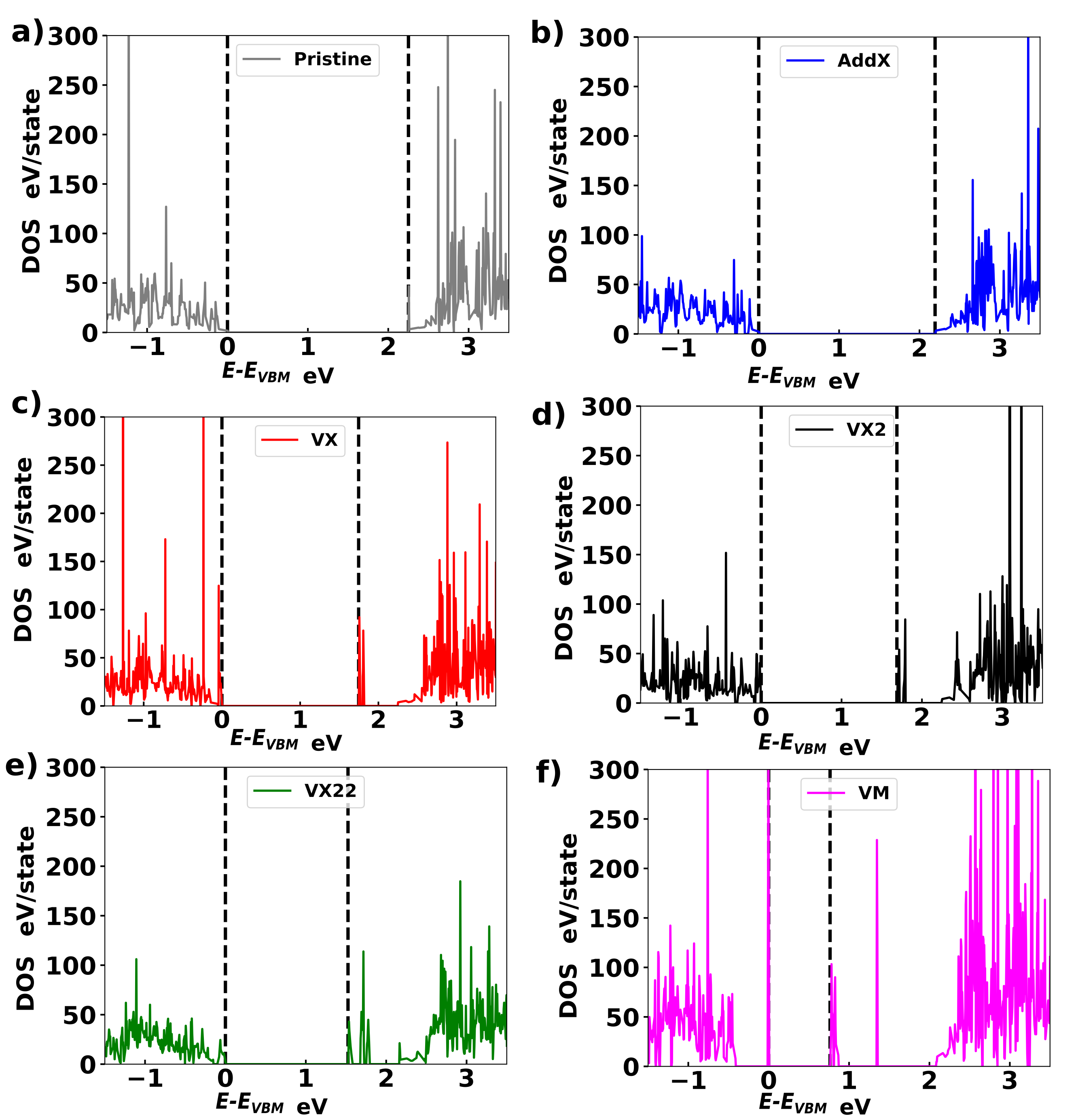}
    \caption{Density of states of MoS$_2$ monolayer for the pristine and point defects under study using HSE06 exchange correlation and SOC.}
      \label{}
\end{figure}
\begin{figure}[h!]
    \centering
 
    \includegraphics[width=1.0\textwidth]{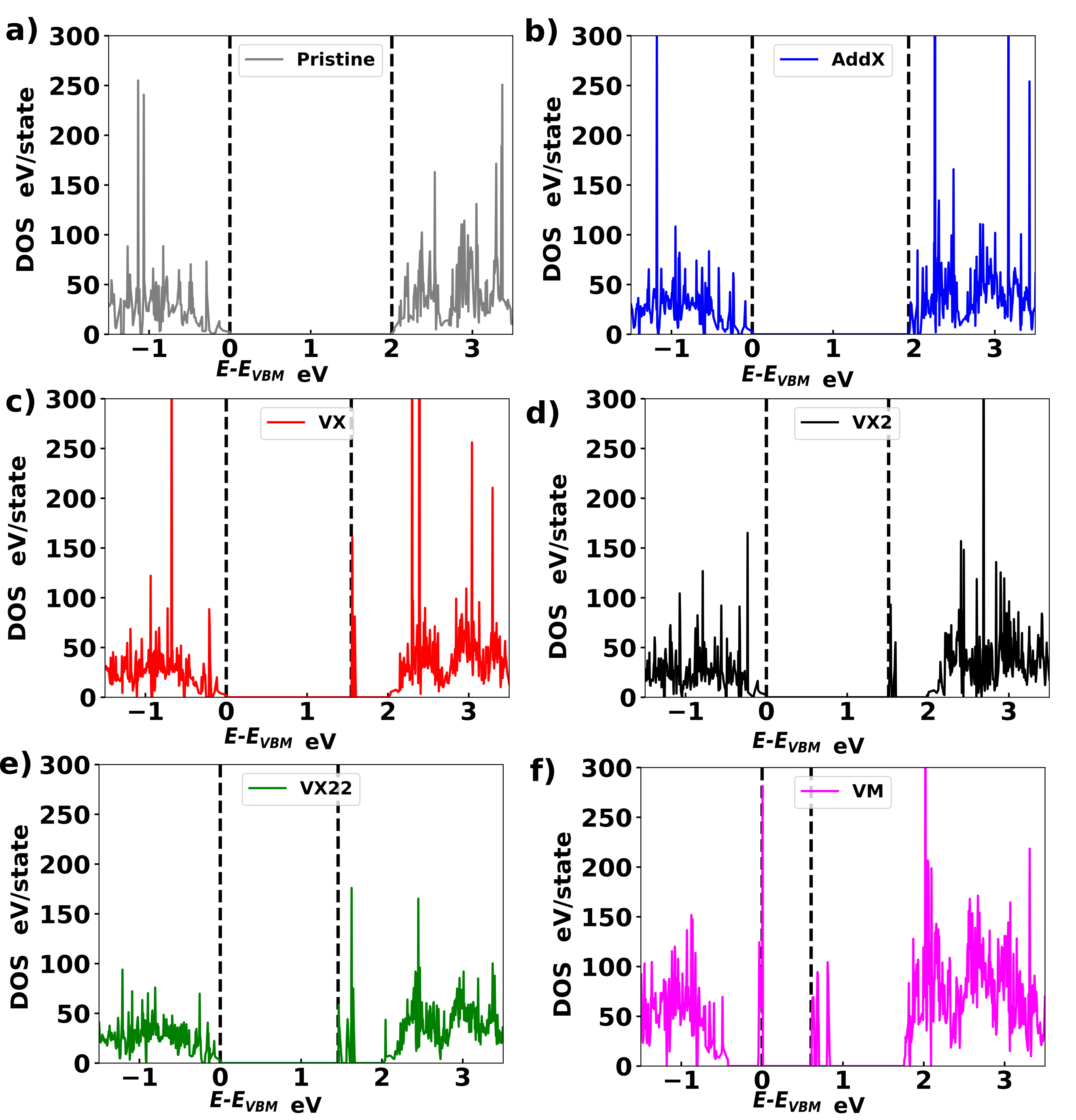}
    \caption{Density of states of MoSe$_2$ monolayer for the pristine and point defects under study using HSE06 exchange correlation and SOC.}
      \label{}
\end{figure}
\begin{figure}[h!]
    \centering
 
    \includegraphics[width=1.0\textwidth]{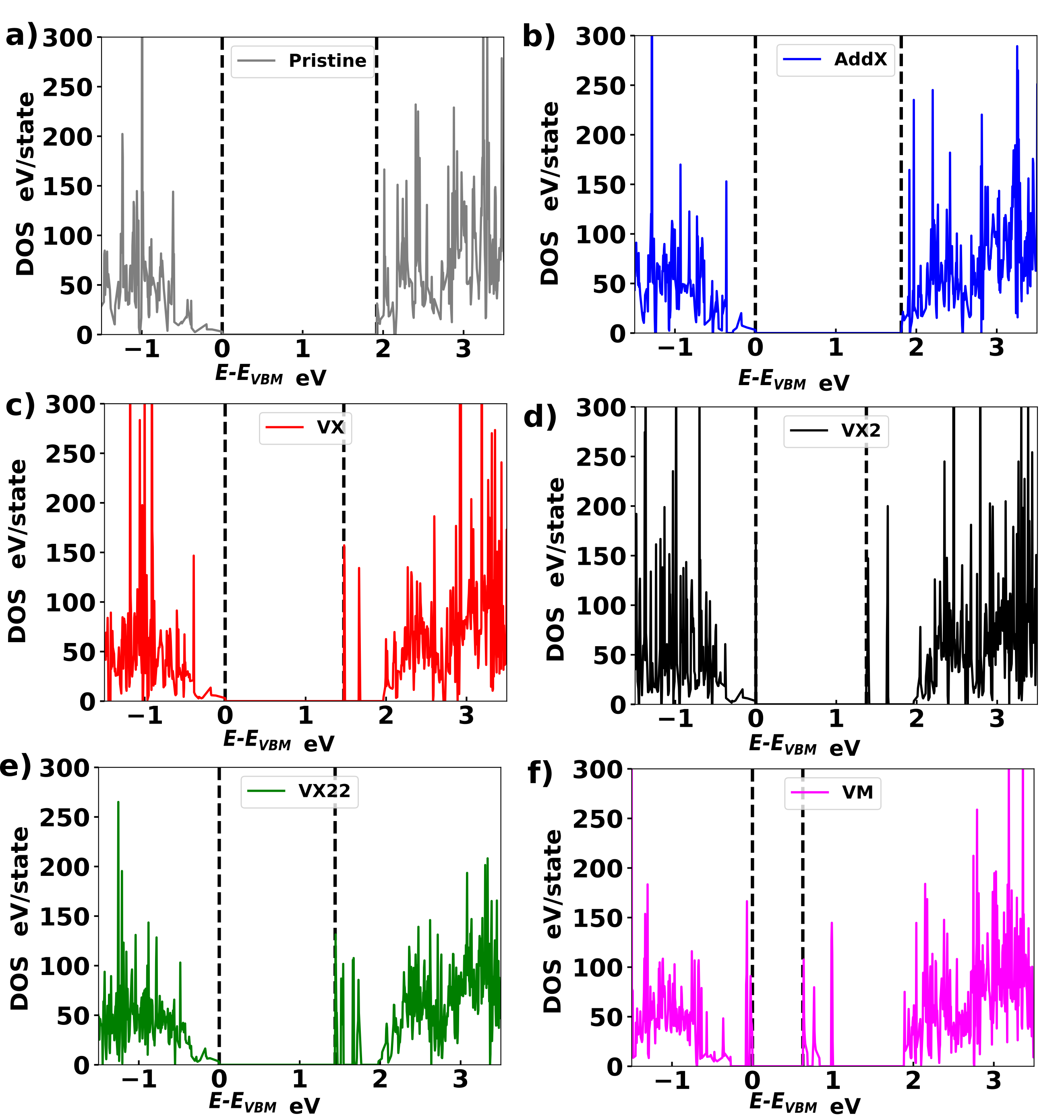}
    \caption{Density of states of WS$_2$ monolayer for the pristine and point defects under study using HSE06 exchange correlation and SOC.}
      \label{}
\end{figure}
\begin{figure}[h!]
    \centering
 
    \includegraphics[width=1.0\textwidth]{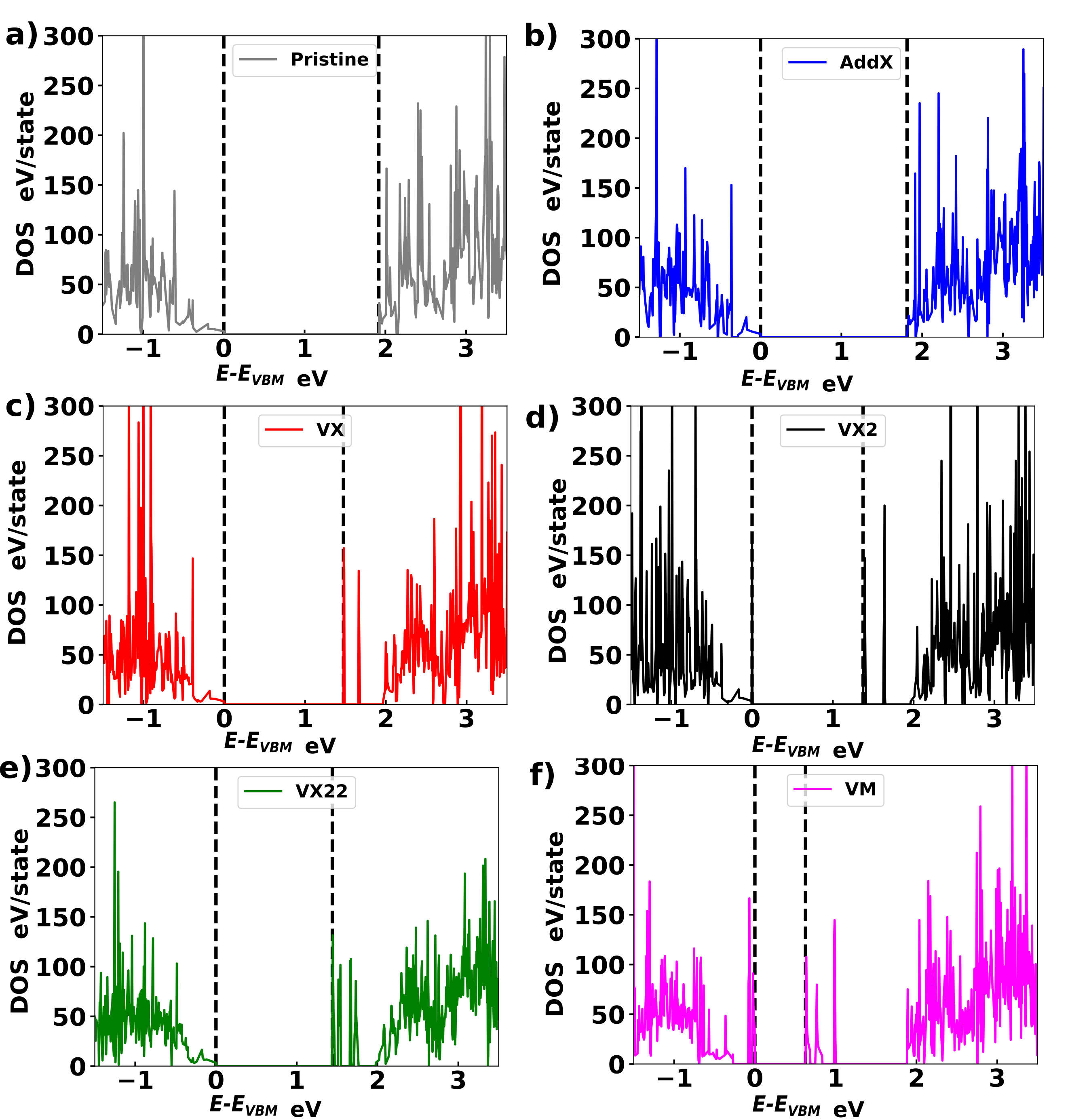}
    \caption{Density of states of WSe$_2$ monolayer for the pristine and point defects under study using HSE06 exchange correlation and SOC.}
      \label{}
\end{figure}

%\clearpage

%\bibliography{paper}% Produces the bibliography via BibTeX.